\documentclass[12pt,leqno, openany, a4paper]{book}
\usepackage{amsmath,amssymb,amsfonts} 
\usepackage{graphics}                 
\usepackage{color}                    
\usepackage{hyperref}                 

\font\ticp=cmcsc15

\makeindex

\def\L{\Lambda}
\newcommand{\bem}{\begin{pmatrix}}
\newcommand{\eem}{\end{pmatrix}}
\def\im{\mathrm{Im}}

\def\be{\begin{equation*}}
\def\ee{\end{equation*}}
\def\beq{\begin{eqnarray*}}
\def\eeq{\end{eqnarray*}}

\def\TrH#1{ {\raise -.5em
                      \hbox{$\buildrel {\textstyle  {\rm Tr } }\over
{\scriptscriptstyle \CH _ {#1}}$}~}}
\def\tr#1{ {\raise -.5em
                      \hbox{$\buildrel {\textstyle  {\rm Tr } }\over
{\scriptscriptstyle{#1}}$}~}}

\renewcommand{\th}{\theta}
\renewcommand{\Im}{\mbox{Im}}
\renewcommand{\Re}{\mbox{Re}}

\newcommand{\IZ}{\mathbb{Z}}

\newcommand{\Tr}{\mbox{Tr}}

\newcommand{\half}{\frac{1}{2}}

\newcommand{\apm}{\alpha'}

\def\s{\sigma}
\def\g{\gamma}
\def\t{\tau}
\def\a{\alpha}
\def\b{\beta}
\def\d{\delta}
\def\m{\mu}

\def\e{\epsilon}

\def\l{{\lambda}}

\def\O{{\Omega}}

\def\G{\Gamma}

\def\CH{{\cal H}}

\def\CM{{\cal M}}

\def\CN{{\mathcal{N}}}

\def\half{{\frac12}}

\def\IC{\relax\hbox{$\inbar\kern-.3em{\rm C}$}}
\def\bZ{{\bf Z}}

\def\bT{{\bf T}}

\def\IC{{\bf C}}

\def\CN{{\cal N}}

\def\bea{\begin{eqnarray}}
\def\eea{\end{eqnarray}}
\def\be{\begin{equation}}
\def\ee{\end{equation}}
\def\ba{\begin{align}}
\def\ea{\end{align}}
\def\bse{\begin{subequations}}
\def\ese{\end{subequations}}
\def\1F1{{}_1\!F_1}
\def\2F0{{}_2\!F_0}

\newenvironment{myenumerate}{
\begin{enumerate}
   \setlength{\itemsep}{1pt}
   \setlength{\parskip}{0pt}
   \setlength{\parsep}{0pt}}{\end{enumerate}}
\newenvironment{myitemize}{
\begin{itemize}
   \setlength{\itemsep}{1pt}
   \setlength{\parskip}{0pt}
   \setlength{\parsep}{0pt}}{\end{itemize}}

\title{\bf Lectures on Quantum Black Holes  }

\author{\ticp Atish Dabholkar$^{1, 2}$\\
\\
\it $^{1}${Theory Division, CERN}\\
\vspace{0.4cm}
\it{PH-TH Case C01600,CH-1211,23 Geneva, Switzerland}\\
\it $^2${Laboratoire de Physique Th\'eorique et Hautes
Energies (LPTHE)}\\
\it{Universit\'e Pierre et Marie Curie-Paris 6; CNRS UMR
7589}\\
\it{Tour 13-14, 5$^{\grave{e}me}$ \'etage, Boite 126, 4 Place
Jussieu} \\
\it {75252 Paris Cedex 05, France}\\
\\
{\ticp Suresh Nampuri}\\
\\
\it{Laboratoire de Physique Th\'eorique}\\
\it{Unit\'e Mixte du CNRS et
    de l'Ecole Normale Sup\'erieure}\\
\it {Ecole Normale Sup\'erieure} \\
\it {24 rue Lhomond, F--75231 Paris Cedex 05, France}\\
}

\date{}
\begin{document}
\maketitle
\pagenumbering{roman}
\tableofcontents
\chapter*{Preface}\normalsize
\pagestyle{plain}
\pagestyle{headings}
\pagenumbering{arabic}

The entropy of black holes supplies us with very useful quantitative information about the fundamental degrees of freedom of quantum gravity. One of the important successes of string theory is that one can explain the thermodynamic entropy of certain supersymmetric black holes as a logarithm of the microscopic degeneracy as required by the Boltzmann relation.  These results imply that  at the quantum level, one should regard  a black hole as an ensemble of quantum states in the  Hilbert space of the theory.

In any consistent quantum theory of gravity such as string theory,  the requirement that  the  thermodynamic entropy must equal the  statistical  entropy of a black hole  is an extremely stringent  theoretical constraint.  This constraint is also \textit{universal} in that  it must hold  in any  `phase' or compactification of the theory that admits a black hole.  It is therefore a  particularly useful guide in our explorations of  string theory in the absence of direct experimental guidance, especially given the fact that we do not know which phase of the  theory might describe the real world. 

Much of the earlier work concerning quantum black holes has been in the limit of large charges when the area of the event horizon is also large. In recent years there has been substantial progress in understanding the entropy of  supersymmetric black holes within string theory going well beyond the large charge limit. It has now become possible to begin exploring finite size effects in perturbation theory in inverse size and even nonperturbatively, with highly nontrivial agreements between thermodynamics and statistical mechanics. Unlike the leading Bekenstein-Hawking entropy which follows from the two-derivative Einstein-Hilbert action, these finite size corrections depend sensitively on `phase' under consideration and contain a wealth of information about the details of compactification as well as  the spectrum of nonperturbative states in the theory. Finite-size corrections are therefore very interesting as  a valuable window into the microscopic degrees of freedom of the theory.

In these notes we describe recent progress in understanding these  finite size corrections to the black hole entropy. To simplify the discussion, we consider the  compactification of the heterotic string on $T^{4} \times T^{2}$ which is dual to  the compactifcation of Type-II string on $K_{3} \times T^{2}$. This leads to a four-dimensional theory with $\CN=4$  supersymmetry and $22$ vector multiplets. Our objective will be to understand the entropy of  half-BPS and quarter-BPS black holes in this theory both from the thermodynamic and statistical view points. A lot is known about generalization of these results to other compactifications. For a review of these generalization and of some of the material covered here  see \cite{Sen:2007qy,Mandal:2010cj}. There has also  been more  progress both in defining the quantum entropy using $AdS/CFT$ correspondence and in computing it using localization.  For a review see \cite{Gomes:2011zz}. We will not discuss these more recent topics here to keep the discussion simple and more accessible.

The organization is as follows. We review aspects of classical and semiclassical black holes in chapters \S\ref{Classical} and  \S{\ref{Semiclassical}}, and elements of string theory in chapter \S\ref{Elements}.  The microscopic counting is then described in chapters \S\ref{Half} and \S\ref{Quarter} and the comparison with macroscopic entropy is discussed  in \S\ref{Quantum}. Relevant mathematical background is covered in \S\ref{Mathematical}. 

These lecture notes are aimed at beginning graduate students but assume some basic background in General Theory of Relativity, Quantum Field Theory, and String Theory. 
A good introductory textbook on general relativity from a modern perspective see \cite{Carroll:2004st}. For a more detailed treatment see \cite{Wald:1984rg} which has become a standard reference among relativists, and \cite{Misner:1974qy} which  remains a classic for various aspects of general relativity. For quantum field theory in curved spacetime see \cite{Birrell:1982ix}. For relevant aspects of string theory see \cite{Green:1987sp, Green:1987mn, Polchinski:1998rq,
Polchinski:1998rr}.  

These notes are based primarily on lectures delivered at the Summer school 2010 in Munich on "Strings and Fundamental Physics."  as well as  at various lectures courses  by AD  on ``Quantum Black Holes''  taught  at the Universit\'e Pierre et Marie Curie, Paris VI together with Ashoke Sen;  at the `School on  D-Brane Instantons, Wall Crossing and Microstate Counting ' at the ICTP Trieste in 2010;  at  the ``School on Black Objects in Supergravity'' at the INFN, Frascati in 2010. Some of the material was  used in earlier  lecture courses by AD at  Shanghai, CERN, Carg\`ese, and Seoul.


\chapter{Classical Black Holes\label{Classical}}

A black hole is at once the most simple and the most complex
object.

It is the most simple in that it is completely specified by its
mass, spin, and charge. This remarkable fact is a consequence of a
the so called `No Hair Theorem'. For an astrophysical object like
the earth, the gravitational field around it depends not only on
its mass  but also on how the mass is distributed  and on the
details of the oblate-ness of the earth and on the shapes of the
valleys and mountains. Not so for a black hole. Once a star
collapses to form a black hole, the gravitational field around it
forgets all details about the star that disappears behind the even
horizon except for its mass, spin, and charge. In this respect, a
black hole is very much like a structure-less elementary particle
such as an electron.

And yet it is the most complex in that it possesses a huge
entropy. In fact the entropy of a solar mass black hole is
enormously bigger than the thermal entropy of the star that might
have collapsed to form it. Entropy gives an account of the number of microscopic states of a system. Hence, the entropy of a black hole  signifies an incredibly complex microstructure. In this respect, a black hole is very unlike an elementary particle.

Understanding the simplicity of a black hole falls in the realm of
classical gravity.  By the early seventies, full fifty years after
Schwarzschild, a reasonably complete understanding of
gravitational collapse and of the properties of an event horizon
was achieved within classical general relativity. The final
formulation began with the  singularity theorems of Penrose, area
theorems of Hawking and culminated in the laws of black hole
mechanics.

Understanding the complex microstructure  of a black hole implied
by its entropy falls in the realm of quantum gravity and is the topic of present lectures. Recent developments have made it clear that a black hole is `simple' not because it is like an elementary particle, but rather because it is like a statistical ensemble. An ensemble is also specified by a few conserved quantum numbers such as energy, spin, and charge. The simplicity of a black hole is  no different than the simplicity that characterizes a thermal ensemble.

To understand the relevant parameters and the geometry of black
holes, let us first consider the Einstein-Maxwell theory described
by the action
\begin{equation}\label{action}
{1 \over 16 \pi G} \int R \sqrt{g} d^4x - {1\over 16 \pi} \int F^2
\sqrt{g} d^4 x,
\end{equation}
where $G$ is Newton's constant, $F_{\mu\nu}$ is the
electro-magnetic field strength, $R$ is the Ricci scalar of the
metric $g_{\mu\nu}$. In our conventions,  the indices $\mu, \nu$
take values $0, 1, 2, 3$ and the metric has signature $(-, +, +,
+)$.

\section{Schwarzschild metric}

Consider the Schwarzschild  metric which is a spherically symmetric,
static solution of the vacuum Einstein equations $R_{\mu\nu}-{1\over
2}g_{\mu\nu} = 0$  that follow from (\ref{action}) when no
electromagnetic fields are excited. This metric is expected to
describe the spacetime outside a gravitationally collapsed
non-spinning star with zero charge. The solution for the line
element is given by
\[
ds^2 \equiv g_{\mu\nu} dx^\mu dx^\nu =  - (1 - {2GM \over r}) dt^2
+ (1 - {2GM \over r})^{-1} dr^2 + r^2 d \Omega^2,
\]
where $t$ is the time, $r$ is the radial coordinate, and $\Omega$
is the solid angle on a 2-sphere. This metric appears to be
singular at $r = 2GM$ because some of its components vanish or
diverge, $g_{00} \to \infty$ and $g_{rr} \to \infty$. As is well
known, this is not a real singularity. This is because the
gravitational tidal forces are finite  or in other words,
components of Riemann tensor are finite in orthonormal
coordinates.  To better understand the nature of this apparent
singularity, let us examine the geometry more closely near $r
=2GM$. The surface $r = 2GM$ is called the `event horizon' of the
Schwarzschild solution. Much of the interesting physics having to
do with the quantum properties of black holes comes from the
region near the event horizon.

To focus on the near horizon geometry in the region $(r - 2GM) \ll
2GM$, let us define $(r - 2GM)= \xi$ , so that when $r \to 2GM$ we
have $ \xi \to 0$. The metric then takes the form
\begin{equation}\label{nearsc}
ds^2 = -{\xi \over 2GM} dt^2 + {2GM \over \xi} (d\xi)^2 + (2GM)^2
d \Omega^2,
\end{equation}
up to corrections that are of order $({1 \over 2GM})$. Introducing
a new coordinate $\rho$,
\[
 \rho^2
= (8GM) \xi  \quad \textrm{so that} \quad d\xi^2 {2GM \over \xi} =
d\rho^2,
\]
the metric takes the form
\begin{equation}\label{metric2}
ds^2 = - {\rho^2 \over 16 G^2M^2} dt^2 + d\rho^2 + (2GM)^2 d
\Omega^2.
\end{equation}
{}From the form of the metric it is clear that $\rho$ measures the
geodesic radial distance. Note  that the geometry factorizes. One
factor is a 2-sphere of radius $2GM$ and the other is the  ($\rho,
t$) space
\begin{equation}\label{sc2}
ds^2_2 = - {\rho^2 \over 16 G^2M^2} dt^2 + d\rho^2.
\end{equation}
We now show that this $1+1$ dimensional spacetime is just a flat
Minkowski space written in funny  coordinates called the Rindler
coordinates.

\section{Rindler coordinates}

To understand Rindler coordinates and their relation to the near
horizon geometry of the black hole,  let us start with  $1+1$
Minkowski space with the usual flat Minkowski metric,
\begin{equation}\label{mink}
ds^2 = -dT^2 + dX^2.
\end{equation}
In light-cone coordinates,
\begin{equation}\label{coord}
U =(T + X ) \quad  V = (T - X),
\end{equation}
the line element takes the form
\begin{equation}\label{ligthcome}
ds^2 = -dU \ dV.
\end{equation}
Now we make a coordinate change
\begin{equation}\label{change}
U =  {1\over \kappa} e^{\kappa u}, \quad V =  - {1\over \kappa}
e^{-\kappa v},
\end{equation}
to introduce the Rindler coordinates $(u, v)$. In these
coordinates the line element takes the form
\begin{equation}\label{rindlermetric0}
ds^2 = -dU \ dV = -e^{\kappa(u-v)} du \ dv.
\end{equation}
Using further coordinate changes
\begin{equation}\label{changecoord}
u = (t + x), \quad v = (t - x), \quad \rho   = {1 \over \kappa}
e^{\kappa x},
\end{equation}
we can write the line element as
\begin{equation}\label{rindler2}
ds^2 = e^{2\kappa x} (-dt^2 + dx^2) = - {\rho^2} \kappa^2 dt^2 +
d\rho^2.
\end{equation}
Comparing (\ref{sc2}) with this Rindler metric, we see that the
$(\rho, t)$ factor of the  Schwarzschild solution  near $r \ \sim
2GM$ looks precisely like Rindler spacetime with metric
\begin{equation}\label{rindlermetric}
ds^2 = -\rho^2 \kappa^2 \ dt^2 + d\rho^2
\end{equation}
with the identification
\[
\kappa = {1\over 4 GM}.
\]
This parameter $\kappa$ is called the surface gravity of the black
hole. For the Schwarzschild solution, one can think of it
heuristically as the Newtonian acceleration $GM/r_H^2$ at the
horizon radius $r_H = 2 GM$. Both these parameters--the surface
gravity $\kappa$ and the horizon radius $r_H$  play an important
role in the thermodynamics of black hole.

This analysis demonstrates that the Schwarzschild spacetime near
$r = 2GM$ is not singular at all. After all it looks exactly like
flat Minkowski space times a sphere of radius $2GM$. So the
curvatures  are inverse powers of the radius of curvature $2GM$
and hence are small for large $2GM$.

\section{Exercises}

\subsubsection{Uniformly accelerated observer and Rindler coordinates}

Consider an astronaut in a spaceship moving with constant acceleration $a$ in Minkowski spactime with Minkowski coordinates $(T, \vec{X})$. 
This means she feels a constant normal reacting from the floor of the spaceship in her rest frame:
\begin{equation}
\frac{d^{2} \vec{X}}{dt^{2}} = \vec{a} \, , \quad \frac{dT}{d\tau} =1
\end{equation}
where   $\tau$ is proper time and $\vec{a}$ is the acceleration 3-vector.
\begin{myenumerate}
\item \textit{Write the equation of motion in a covariant form and show that her 4-velocity $u^{\m}:=\frac{dX^{\m}}{d\t}$ is timelike whereas her 4-acceleration $a^{\m}$ is spacelike. }
\item \textit{Show that if she is moving along the $x$ direction, then her trajectory is of the form
\begin{equation}
T = \frac{1}{a} \sinh(a\t) \, , \quad  X = \frac{1}{a} \cosh(a\t)
\end{equation}
}
which is a hyperboloid. Find the acceleration 4-vector. 
\item \textit{Show that it is natural for her to use her proper time as the time coordinate and introduce  a coordinate frame of  a family of observers with
\begin{equation}
T =\zeta \, \sinh(a\eta) \, , \quad  X = \zeta \,\cosh(a\eta) \, .
\end{equation}}
\end{myenumerate}
By examining the metric, show that $v =\eta -\zeta$ and $u = \eta +\zeta$ are precisely the Rindler coordinates introduced earlier with the acceleration parameter $a$ identified with the surface gravity $\kappa$.

\section{Kruskal extension}

One important fact to note about the Rindler metric is  that the
coordinates $u,v$ do not cover all of Minkowski space because even
when the vary over the full range
\[
{-\infty \leq u \leq \infty, \quad  -\infty \leq v \leq \infty }
\]
the Minkowski coordinate vary only over the quadrant
\begin{equation}\label{minkco}
0 \leq U \leq \infty,  \quad   -\infty < V \leq 0.
\end{equation}
If we had written the flat metric in these `bad', `Rindler-like'
coordinates, we would find a fake singularity at $\rho =0$  where
the metric appears to become singular. But we can discover the
`good', Minkowski-like coordinates $U$ and $V$ and extend them to
run from $-\infty$ to $\infty$ to see the entire spacetime.

Since the Schwarzschild solution in the usual $(r, t)$
Schwarzschild coordinates  near $r = 2GM$ looks  exactly like
Minkowski space in Rindler coordinates, it suggests that we must
extend it in properly chosen `good' coordinates. As we have seen,
the  `good' coordinates near $r=2GM$ are related to the
Schwarzschild coordinates in exactly the same way as the Minkowski
coordinates are related the Rindler coordinates.

In fact one can choose  `good' coordinates over the entire
Schwarzschild spacetime. These `good' coordinates are called the
Kruskal coordinates.  To obtain the Kruskal coordinates, first
introduce the `tortoise coordinate'
\begin{equation}\label{tort}
r^\ast= r + 2GM \log \left({r-2GM \over 2GM}\right).
\end{equation}
In the $(r^*, t)$  coordinates, the metric is conformally flat,
\emph{i.e.}, flat up to rescaling
\begin{equation}\label{sc3}
ds^2 = (1 - {2GM \over r}) (-dt^2 + dr^{\ast 2}).
\end{equation}

Near the horizon the coordinate $r^*$ is similar to the coordinate
$x$ in (\ref{rindler2}) and hence  $u = t + r^\ast$ and $v = t -
r^\ast$ are like the Rindler $(u,v)$ coordinates. This suggests
that we define $U,V$ coordinates as in (\ref{change}) with $\kappa
= 1/4GM$. In these coordinates the metric takes the form
\begin{equation}\label{kruskal}
ds^2 = -e^{-(u-v)\kappa} dU \ dV = - {2GM \over r}e^{-r/2GM} dU \
dV
\end{equation}
We now see that the  Schwarzschild coordinates cover only a part
of spacetime because they cover only a  part of the range of the
Kruskal coordinates. To see the entire spacetime, we must extend
the Kruskal coordinates to run from $-\infty$ to $\infty$. This
extension of the Schwarzschild solution is known as the Kruskal
extension.

Note that now the metric is perfectly regular at $r=2GM$ which is
the surface $UV =0$ and there is no singularity there. There is,
however, a real singularity at $r=0$ which cannot be removed by a
coordinate change because physical tidal forces become infinite.
Spacetime stops at $r=0$ and at present we do not know how to
describe physics near this region.

\section{Event horizon}

We have seen that $r= 2GM$ is not a real singularity but a mere
coordinate singularity which can be removed by a proper choice of
coordinates. Thus, locally there is nothing special about the
surface $r=2GM$. However, globally, in terms of the causal
structure of spacetime, it is a special surface and is  called the
`event horizon'. An event horizon is a boundary of region in
spacetime from behind which no causal signals can reach the
observers sitting far away at infinity.

To see the causal structure of the event horizon,  note that in
the metric (\ref{rindler2}) near the horizon, the constant radius
surfaces are determined by
\begin{equation}\label{hyper}
\rho^2 = {1\over \kappa^2} e^{2\kappa x} = {1\over \kappa^2}
e^{\kappa u} e^{-\kappa v} = - UV =  \textrm{constant}
\end{equation}
These surfaces are thus hyperbolas. The Schwarzschild metric is
such that at $r \gg 2GM$ and observer who wants to remain at a
fixed radial distance $r$ = constant is almost like an  inertial,
freely falling observers in flat space. Her trajectory is
time-like and is a straight line going upwards on a spacetime
diagram. Near $r = 2GM$, on the other hand, the constant $r$ lines
are hyperbolas which are the trajectories of observers in uniform
acceleration.

To understand the trajectories of observers at radius $r
> 2GM$, note that to  stay at a fixed radial distance $r$ from a
black hole, the observer must boost the rockets to overcome
gravity. Far away, the required acceleration is negligible and the
observers are almost freely falling. But near $r=2GM$ the
acceleration is substantial and the observers are not freely
falling. In  fact at $r=2GM$, these trajectories are light like.
This means that a fiducial observer who wishes to stay at $r=2GM$
has to move at the speed of light with respect to the freely
falling observer. This can be achieved only with infinitely large
acceleration. This unphysical acceleration is the origin of the
coordinate singularity of the Schwarzschild coordinate system.

In summary, the  surface defined by $r = \textrm{contant}$ is
timelike for $r > 2GM$, spacelike for $r <2GM$, and light-like or
null at $r=2GM$.

In Kruskal coordinates, at $r=2GM$,  we have $UV = 0$ which can be
satisfied in two ways. Either $V=0$, which defines the `future
event horizon', or $U=0$, which defines the `past event horizon'.
The future event horizon is a one-way surface that signals can be
sent into but cannot come out of. The region bounded by the event
horizon is then a black hole. It is literally a hole in spacetime
which is black because no light can come out of it. Heuristically,
a black hole is black because even light cannot escape its strong
gravitation pull. Our analysis of the metric makes this notion
more precise. Once an observer falls inside the black hole she can
never come out because to do so she will have to travel faster
than the speed of light.

As we have noted already $r=0$ is a real singularity that is
inside the event horizon.  Since it is a spacelike surface, once a
observer falls insider the event horizon, she is sure to  meet the
singularity at $r=0$ sometime in future no matter how much she
boosts the rockets.

 In our example of the Schwarzschild black
hole, the event horizon is  static because it is defined as a constant $r$ hypersurface
$r=2GM$ which does not change with time. More precisely, the
time-like Killing vector $\partial \over
\partial t$ leaves it invariant. It is at the same time null
because $g^{rr}$ vanishes at $r=2GM$ so that the norm of the 1-form $dr$ vanishes. 
In general, as for a spinning Kerr-Newman black hole,  the horizon is not static but only stationary (because of the uniform rotation) and null.

In summary, an event horizon is a surface that is simultaneously \emph{stationary} and \emph{null}, which causally separates the inside
and the outside of a black hole. For a  discussion of the notion of an event horizon in greater generality see \cite{Carroll:2004st, Wald:1984rg}.

\section{Black hole parameters}

{}From our discussion of the Schwarzschild black hole we are ready
to abstract some important general concepts that are useful in
describing the physics of more general black holes.

To begin with, a  \textsl{black hole} is an asymptotically flat
spacetime that contains a region which is not in the backward
lightcone of future timelike infinity.  The boundary of such a
region is a stationary null surface call the \textsl{event
horizon}. The fixed $t$ slice of the event horizon is a two
sphere.

There are a number of important parameters of the black hole. We
have introduced these in the context of Schwarzschild black holes.
For a general black holes their actual values are different but
for all black holes, these parameters govern the thermodynamics of
black holes.

\begin{enumerate}
\item The radius of the event horizon $r_H$ is the radius of the
two sphere. For a Schwarzschild black hole, we have $r_H = 2GM$.

\item The area of the event horizon $A_H$ is  given by $4\pi
r_H^2$. For a Schwarzschild black hole, we have $A_H = 16 \pi G^2
M^2$.

\item The surface gravity is the parameter $\kappa$ that we
encountered earlier. As we have seen, for  a Schwarzschild black
hole, $\kappa= 1/4GM$.
\end{enumerate}

\section{Laws of black hole mechanics}

One of the remarkable properties of black holes is that one can
derive a set of laws of black hole mechanics which bear a very
close resemblance to the  laws of thermodynamics. This is quite
surprising because \emph{a priori} there is no reason to expect
that the spacetime geometry of black holes has anything to do with
thermal physics.

\begin{myenumerate}
\item[{(0)}] Zeroth Law: In thermal physics, the zeroth law states
that the temperature $T$ of a body at thermal equilibrium is
constant throughout the body. Otherwise heat will flow from hot
spots to the cold spots. Correspondingly for stationary black holes one can
show that surface gravity $\kappa$ is constant on the event
horizon. This is obvious for spherically symmetric horizons but is
true also more generally for non-spherical horizons of spinning
black holes.

\item[{(1)}] First Law: Energy is conserved, $dE = TdS + \mu dQ +
\Omega dJ$,  where E is the energy, Q is the charge with chemical
potential $\mu$ and  $J$ is the spin with chemical potential
$\Omega$. Correspondingly for black holes, one has $dM = {\kappa
\over 8\pi G} dA + \mu dQ + \Omega dJ$. For a Schwarzschild black
hole we have $\mu = \Omega = 0$ because there is no charge or
spin.

\item[{(2)}] Second Law: In a physical process the total entropy
$S$ never decreases, $\Delta S \geq 0$. Correspondingly for black
holes one can prove the area theorem that the net area in any process never
decreases,  $\Delta A \geq 0$. For example,  two Schwarzschild
black holes with masses $M_1$ and $M_2$  can coalesce to form a
bigger black hole of mass $M$. This is consistent with the area
theorem, since the area is proportional to the square of the mass,
and $(M_1 + M_2)^2 \geq M^2_1 + M^2_2$. The opposite process where
a bigger black hole fragments is however disallowed by this  law.
\end{myenumerate}

Thus the laws of black hole mechanics, crystallized by Bardeen,
Carter, Hawking, and other bears  a striking resemblance with the
three laws of thermodynamics for a body in thermal equilibrium.
We summarize these results below in Table\eqref{blackholelaws} for a black hole of 
mass $M$,  spin $J$,  and charge $Q$.

\begin{table}[h]
\caption{\small{Laws of Black Hole Mechanics}}
\vspace{7mm}
\centering
\begin{tabular}{|c|c|}
\hline
\textbf{Laws of Thermodynamics} & \textbf{Laws of Black Hole Mechanics}\\
 & \\
 \hline Temperature is constant
 & Surface gravity is constant
 \\
 throughout a body at equilibrium. &  on the event horizon.\\
 T=
constant. & $\kappa$ =constant.\\
 \hline
Energy is conserved.
& Energy is conserved. \\
$dE = T dS + \mu dQ + \Omega dJ. $& $dM = \frac{\kappa}{8\pi} dA + \mu dQ + \Omega dJ . $\\
 \hline
 Entropy never decrease.  & Area never decreases.\\
 $\Delta S \geq 0$. & $ \Delta A \geq 0 $. \\
 \hline
\end{tabular}
\label{blackholelaws}
\end{table}
Here $A$ is the area of the horizon,  and $\kappa$ is the surface gravity which can be thought of
roughly as the acceleration at the horizon, $\mu$ is the chemical potential conjugate to $Q$, and $\Omega$ is the angular speed conjugate to $J$. 

We will see that this  formal analogy between the laws of black hole mechanics and thermodynamics  is actually much more than an analogy.
Bekenstein and Hawking discovered that there is a deep connection
between black hole geometry, thermodynamics and quantum mechanics.
Quantum mechanically, a black hole is not quite black.

\section{Historical aside}

Apart from its physical significance, the entropy of a black hole
makes for a fascinating study in the history of science. It is one
of the very rare examples where a scientific idea has gestated and
evolved over several decades into an important conceptual and
quantitative tool almost entirely on the strength of theoretical
considerations. That we can proceed so far with any confidence at
all with very little guidance from experiment is indicative of the
robustness of the basic tenets of physics. It is therefore
worthwhile to place black holes and their entropy in a broader
context before coming to the more recent results pertaining to the
quantum aspects of black holes within string theory.

A black hole is now so much a part of our vocabulary that it can
be difficult to appreciate the initial intellectual opposition  to
the idea of `gravitational collapse' of a star and of a `black
hole' of nothingness in spacetime by several leading physicists,
including Einstein himself.

To quote the relativist Werner Israel ,

``\textsl{ There is a curious parallel between the histories of black
holes and continental drift. Evidence for both was already
non-ignorable by 1916, but both ideas were stopped in their tracks
for half a century by a resistance bordering on the irrational.}''

On January 16, 1916, barely two  months after Einstein had
published the final form of his field equations for gravitation
\cite{Einstein:1915ih}, he presented a paper to the Prussian
Academy on behalf of Karl Schwarzschild
\cite{Schwarzschild:1916sh}, who was then fighting a war on the
Russian front. Schwarzschild had found a spherically symmetric,
static and exact solution of the full nonlinear equations of
Einstein without any matter present.

The Schwarzschild solution  was immediately accepted as the
correct description within general relativity of the gravitational
field outside a spherical mass. It would be the correct
approximate description of the field around a star such as our
sun. But something much more bizzare was implied by the solution.
For an object of mass M, the solution appeared to become singular
at a radius $R = 2 GM/c^2$. For our sun, for example, this radius,
now known as the Schwarzschild radius, would be  about three
kilometers. Now, as long the physical radius of the sun is bigger
than three kilometers, the `Schwarzschild's singularity' is of no
concern because inside the sun the Schwarzschild solution is  not
applicable as there is matter present. But what if the entire mass
of the sun was concentrated in a sphere of radius smaller than
three kilometers? One would then have to face up to this
singularity.

Einstein's reaction to the `Schwarzschild singularity' was to seek
arguments that would make such a singularity inadmissible.
Clearly, he believed,  a physical theory could not tolerate such
singularities. This drove his to write as late as 1939, in a
published paper,

``\textsl{The essential result of this investigation is a clear
understanding as to why the `Schwarzschild singularities' do not
exist in physical reality.}''

This conclusion was however based on an incorrect argument.
Einstein was not alone in this rejection of the unpalatable idea
of a total gravitational collapse of a physical system. In the
same year, in an astronomy conference in Paris, Eddington, one of
the leading astronomers of the time, rubbished the work of
Chandrasekhar who had concluded from his study of white dwarfs, a
work that was to earn him the Nobel prize later, that a large
enough star could collapse.

It is interesting that  Einstein's paper on the inadmissibility of
the Schwarzschild singularity appeared only two months before
Oppenheimer and Snyder published their definitive work on stellar
collapse with an abstract that read,

``\textsl{When all thermonuclear sources of energy are exhausted, a
sufficiently heavy star will collapse.''}

Once a sufficiently big star ran out of its nuclear fuel, then
there was nothing to stop the inexorable inward pull of gravity.
The possibility of stellar collapse meant that a star could be
compressed in a region smaller than its Schwarzschild radius and
the  `Schwarzschild singularity' could no longer be wished away as
Einstein had desired. Indeed it was essential to understand what
it means to understand the final state of the star. 

It is thus useful to keep in mind what seems now like a mere change of coordinates was at one point a matter of raging intellectual debate.

\chapter{Semiclassical Black Holes \label{Semiclassical}}

 In the semiclassical treatment  of a black hole, we treat the spacetime geometry of the black hole classically but  treat  various fields such as the electromagnetic field in this fixed spacetime background quantum mechanically. This semiclassical inclusion of quantum effects already reveals a deep and unexpected connection between the spacetime geometry of a black hole  and thermodynamics.
 
\section{Hawking temperature}

Bekenstein asked a simple-minded but incisive question. If nothing
can come out of a black hole, then a black hole will violate the
second law of thermodynamics. If we throw a bucket of hot water
into a black hole then the net entropy of the world outside would
seem to decrease.  Do we have to give up  the second law of
thermodynamics in the presence of black holes?

Note that the energy of the bucket is also lost to the outside
world but that does not violate the first law of thermodynamics
because the black hole  carries mass or equivalently energy. So
when the bucket falls in, the mass of the black hole goes up
accordingly to conserve energy. This suggests that one can save
the second law of thermodynamics if somehow the black hole also
has entropy. Following this reasoning and noting the formal
analogy between the area of the black hole and entropy discussed
in the previous section, Bekenstein proposed that a black hole
must have entropy proportional to its area \cite{Bekenstein:1973ur}.

This way of saving the second law is however in contradiction with
the classical properties of a black hole because if a black hole
has energy $E$ and entropy $S$, then it must also have temperature
$T$ given by
\[
{1 \over T} = {\partial S \over \partial E}.
\]
For example, for  a Schwarzschild black hole, the area and the
entropy scales as $ S \sim M^2$. Therefore, one would expect
inverse temperature that scales as $M$
\begin{equation}\label{temperature}
{1 \over T} = {\partial S \over \partial M} \sim {\partial M^2
\over
\partial M} \sim M.
\end{equation}
Now, if the black hole has temperature then like any hot body, it
must radiate. For a classical black hole, by its very nature, this
is impossible. 

Hawking showed that after including quantum
effects, however, it is possible for a black hole to radiate \cite{Hawking:1974sw}. In a
quantum theory, particle-antiparticle are constantly being created
and annihilated even in vacuum. Near the horizon, an antiparticle
can fall  in once in a while and the particle can escapes to
infinity. In fact, Hawking's calculation showed that the spectrum
emitted by the black hole is  precisely thermal with temperature
$T = {\hbar\kappa \over 2\pi} = {\hbar \over 8\pi GM}$. With this
precise relation between the temperature and surface gravity the
laws of black hole mechanics discussed in the earlier section
become identical to the laws of thermodynamics.  Using the formula
for the Hawking temperature and the first law of thermodynamics
\[
dM = TdS = {\kappa \hbar \over 8\pi G\hbar} dA,
\]
one can then deduce the precise relation between entropy and the
area of the black hole:
\[
 S = {A c^3 \over 4G\hbar} \, .
\]

Before discussing the entropy of a black hole, let us derive  the
Hawking temperature in a somewhat heuristic way using a Euclidean
continuation of the near horizon geometry. In quantum mechanics,
for a system with Hamiltonian $H$, the thermal partition function
is
\begin{equation}\label{thermal}
Z = \textrm{Tr} e^{-\beta \hat H},
\end{equation}
where $\beta$ is the inverse temperature. This is related to the
time evolution operator $e^{-itH/\hbar}$ by a Euclidean analytic
continuation $t = -i\tau$ if we identify $\tau = \beta \hbar$. Let
us consider a single scalar degree of freedom $\Phi$, then one can
write the trace as
\[ \textrm{Tr} e^{-\tau \hat H/\hbar}
 = \int d \phi < \phi | e^{-\tau_E \hat
H/\hbar} |\phi>
\]
and use the usual path integral representation for the propagator
to find
\[
\textrm{Tr} e^{-\tau \hat H/\hbar} =  \int d \phi \int D\Phi
e^{-S_E[\Phi]}.
\]
Here $S_E[\Phi]$ is the Euclidean action over periodic field
configurations that satisfy the boundary condition
\[
\Phi (\beta\hbar) = \Phi (0) = \phi.
\]
This gives the relation between the periodicity in Euclidean time
and the inverse temperature,
\begin{equation}\label{temp}
\beta \hbar = \tau \quad \textrm{or} \quad T = {\hbar \over \tau}.
\end{equation}
Let us now look at the Euclidean Schwarzschild metric by
substituting $t = -i t_E$. Near the horizon the line element
(\ref{rindler2}) looks like
\[
ds^2 = \rho^2 \kappa^2 dt_E^2 + d\rho^2.
\]
If we now write $\kappa t_E = \theta$, then this metric is just
the flat two-dimensional Euclidean metric written in polar
coordinates provided the angular variable $\theta$ has the correct
periodicity $0 < \theta < 2\pi$. If the periodicity is different,
then the geometry would have a conical singularity at $\rho =0$.
This implies  that Euclidean time $t_E$ has periodicity $\tau =
{2\pi \over \kappa}$. Note that far away from the black hole at
asymptotic infinity the Euclidean metric is flat and goes as $ds^2
= d\t_E^2 + dr^2$. With periodically identified Euclidean time,
$t_E \sim t_E + \tau$, it looks like a cylinder. Near the horizon
at $\rho=0$ it is nonsingular and looks like flat space in polar
coordinates for this correct periodicity. The full Euclidean
geometry thus looks like a cigar.  The tip of the cigar is at
$\rho =0$  and the geometry is asymptotically cylindrical far away
from the tip.

Using the relation between Euclidean periodicity and temperature,
we then conclude that Hawking temperature of the black hole is
\begin{equation}\label{hawktemp}
T = {\hbar \kappa \over 2 \pi}.
\end{equation}

\section{Bekenstein-Hawking entropy}

Even though we have ``derived'' the temperature and the entropy in
the context of Schwarzschild black hole, this beautiful relation
between area and entropy is true quite generally essentially
because the near horizon geometry is always Rindler-like. For
\emph{all} black holes with charge, spin and in number of
dimensions, the Hawking temperature and the entropy are given in
terms of the surface gravity and horizon area by the formulae
\[
T_H = {\hbar \kappa \over 2\pi}, \quad S = {A \over 4G \hbar}.
\]
This is a remarkable relation between the thermodynamic properties
of a black hole on one hand and its geometric properties on the
other.

The fundamental significance of entropy stems from the fact that
even though it  is a quantity defined in terms of gross
thermodynamic properties, it contains nontrivial information about the
\textit{microscopic} structure of the theory through Boltzmann
relation
\[
S = k \log (d),
\]
where $d$ is the the degeneracy or the total number of microstates of the system of
for a given energy, and $k$ is Boltzmann constant. Entropy is not
a kinematic quantity like energy or momentum but  rather contains
information about the total number microscopic degrees of freedom
of the system. Because of the Boltzmann relation, one can learn a great deal
about the microscopic properties of a system from its
thermodynamics properties.

The Bekenstein-Hawking entropy behaves in every other respect like
the ordinary thermodynamic entropy. It is therefore natural to ask
what microstates might account for it. Since the entropy formula
is given by this  beautiful, general form
\[
S = {Ac^3 \over 4G\hbar} ,
\]
that involves all three fundamental dimensionful constants of
nature, it is a valuable piece of information about the degrees of
freedom of a quantum theory of gravity.

\section{Exercises}

\subsubsection{Reissner-Nordstr\"om (RN) black hole}

The most general static, spherically symmetric, charged  solution
of the Einstein-Maxwell theory (\ref{action}) gives the
Reissner-Nordstr\"om (RN) black hole. In what follows  we choose
units so that $G = \hbar =1$. The line element is given by
\begin{equation}\label{rn}
ds^2 = -\left(1 - {2M \over r} + {Q^2 \over r^2}\right) dt^2 +
\left(1 - {2M \over r} + {Q^2 \over r^2}\right)^{-1} dr^2 + r^2 d
\Omega^2,
\end{equation}
and the electromagnetic field strength by
\[
F_{tr} = Q/r^2.
\]
The parameter $Q$ is the charge of the black hole and $M$ is the mass. For $Q=0$ this reduces to  the Schwarzschild black hole.

{}From the metric \eqref{rn} we see that the event horizon for this solution is located at where
$g^{rr} =0$, or
\[
1 - {2M \over r} + {Q^2 \over r^2} = 0.
\]
Since this is a quadratic equation in $r$,
\[
r^2 - 2 QMr + Q^2 = 0,
\]
it has two solutions.

\[
r_\pm = M \pm \sqrt{M^2 - Q^2}.
\]
Thus, $r_+$ defines the outer horizon of the black hole and $r_-$
defines the inner horizon of the black hole. The area of the black
hole is  $4\pi r_+^2$.
\begin{myenumerate}
\item \textit{Identify the horizon for this metric and examine the near horizon geometry to show that it has two-dimensional Rindler spacetime as a factor. }

\item \textit{Using the relation to the Rindler geometry determine the surface gravity $\kappa$ as  for the Schwarzschild black hole and thereby determine the temperature and entropy of the black hole.
 \begin{eqnarray}\label{temprn} \nonumber
  T &=&{\kappa \hbar \over 2\pi} =
  {\sqrt{M^2 - Q^2} \over 2\pi (2 M (M + \sqrt{M^2 - Q^2}) - Q^2)} \\
\nonumber   S &=& \pi r^2_+ = \pi (M + \sqrt{M^2 - Q^2})^2.
\end{eqnarray}
Recover the formulae for Schwarzschild black hole in
the limit $Q = 0$.
}

\item \textit{Show that in the extremal limit $M \rightarrow Q$ the temperature vanishes but the entropy has a nonzero limit. Show that for the extremal Reissner-Nordstr\"om black hole the near horizon geometry is of the form $AdS_{2}\times S^{2}$.}
\end{myenumerate}

\section{ Bekenstein-Hawking-Wald entropy \label{Wald}}

In our discussion of Bekenstein-Hawking entropy of a black hole,
the Hawking temperature could be deduced from surface gravity or
alternatively the periodicity of the Euclidean time in the black
hole solution. These are  geometric asymptotic properties of the
black hole solution. However, to find the entropy we needed to use
the first law of black hole mechanics which was derived in the
context of Einstein-Hilbert action
\[
{1 \over 16\pi} \int R \sqrt{g} d^4x.
\]

Generically in string theory, we expect corrections (both in
$\alpha'$ and $g_s$) to the effective action that has higher
derivative terms involving Riemann tensor and other fields.
\[
I = {1 \over 16\pi} \int (R + R^2 + R^4 F^4 + \cdots).
\]
How do the laws of black hole thermodynamics get modified?

Wald derived the first law of thermodynamics in the presence of
higher derivative terms in the action \cite{Wald:1993nt, Iyer:1994ys, Jacobson:1994qe}. This generalization implies
an elegant formal expression for the entropy $S$ given a general
action $I$ including higher derivatives
\[
S = 2\pi \int_{\rho^2} {\delta I \over \delta
R_{\mu\nu\alpha\beta}} \epsilon^{\mu\a} \epsilon^{\nu\beta}
\sqrt{h} d^2 \Omega,
\]
where $\epsilon^{\mu\nu}$ is the binormal to the horizon, $h$ the
induced metric on the horizon, and the variation of the action
with respect to $R_{\mu\nu\alpha\beta}$ is to be carried out
regarding the Riemann tensor as formally independent of the metric
$g_{\mu\nu}$.

As an example, let us consider the Schwarzschild solution of the
Einstein Hilbert action. In this case, the event horizon is  $S^2$
which has two  normal directions along $r$ and $t$. We can
construct an antisymmetric 2-tensor $\epsilon_{\mu\nu}$ along
these directions so that $\epsilon_{rt} = \epsilon_{tr} =-1$. \beq
\mathcal{L} = {1\over 16\pi} R_{\mu\nu\alpha\beta} g^{\nu\alpha}
g^{\mu\beta}, \quad {\partial \mathcal{ L} \over
\partial R_{\mu\nu\alpha\beta}} ={1 \over 16\pi} {1\over2}
(g^{\mu\alpha} g^{\nu\beta} - g^{\nu\alpha} g^{\mu\beta}) \eeq
Then the Wald entropy is given by
\beq
S &=& {1\over8} \int
{1\over2} (g^{\mu\alpha} g^{\nu\beta} - g^{\nu\alpha}
g^{\mu\beta}) (\epsilon_{\mu\nu} \epsilon_{\alpha\beta})
\sqrt{h} d^2\Omega \nonumber \\
&=& {1\over8} \int g^{tt} g^{rr} \cdot 2 = {1\over4} \int_{S^2}
\sqrt{h} d^2\Omega = {A_H \over 4}, \nonumber \eeq
giving us the Bekenstein-Hawking formula as expected.

\section{Extremal Black Holes}

For a physically sensible definition of temperature and entropy in
(\ref{temprn}) the mass must satisfy the bound $M^2 \geq Q^2$.
Something special happens when this bound is saturated and $M =
|Q| $. In this case $r_+ = r_- = |Q|$ and   the two horizons
coincide. We choose $Q$ to be positive. The solution (\ref{rn})
then takes the form,
\begin{equation}\label{rnextermal}
ds^2 = - (1 - Q/r)^2 dt^2 + {dr^2 \over (1-Q/r)^2} + r^2 d
\Omega^2,
\end{equation}
with a horizon at $r = Q$. In this extremal limit (\ref{temprn}),
we see that the temperature of the black hole goes to zero and it
stops radiating but nevertheless its entropy has a finite limit
given by $S \rightarrow \pi Q^2$. When the temperature goes to
zero, thermodynamics does not really make sense but we can use
this limiting entropy as the definition of the zero temperature
entropy.

For extremal black holes it is sometimes  more convenient to use isotropic
coordinates in which the line element takes the form
\[
ds^2 = H^{-2} (\vec x) dt^2 + H^2 (\vec x) d\vec x^2
\]
where $d\vec x^2$ is the flat Euclidean line element
$\delta_{ij}dx^i dx^j$ and $H(\vec x)$ is a harmonic function of
the flat Laplacian
\[
\delta^{ij} {\partial \over \partial x^i} {\partial \over
\partial x^j}.
\]
The extremal Reissner-Nordstr\"om solution is obtained by choosing
\[
H (\vec x) = \left(1 + {Q \over \rho}\right),
\]
and the field strength is given by $F_{0i} = \partial_i H(\vec
x)$.

One can in fact write a multi-centered Reissner-Nordstr\"om
solution by choosing a more general harmonic function
\begin{equation}\label{harmonic}
H = 1 + \sum^N_{i=1} {Q_i \over |\vec x - \vec xi|}.
\end{equation}
The total mass $M$ equals the total charge $Q$ and is  given
additively
\begin{equation}\label{total}
Q = \sum Q_i.
\end{equation}
The solution is static because the electrostatic repulsion between
different centers balances the gravitational attraction between them.

Note that the coordinate $\rho$ in the isotropic coordinates should
not be confused with the coordinate $r$ in the spherical
coordinates. In the isotropic coordinates the line-element is
\[
ds^2 = - \left( 1 + {Q \over \rho}\right)^2 dt^2 +  (1+{Q\over
\rho})^{-2} (d\rho^2 + \rho^2 d \Omega^2),
\]
and the horizon occurs at $\rho = 0$. Contrast this with the metric
in the spherical coordinates (\ref{rnextermal}) that has the
horizon at $r= Q$.  The near horizon geometry is quite different
from that of the Schwarzschild black hole. The line element is
\beq
ds^2&=& - {\rho^2 \over Q^2} dt^2 + {Q^2 \over \rho^2} (d\rho^2 + \rho^2 d \Omega^2) \\
&=& (-{\rho^2 \over Q^2} dt^2 + {Q^2 \over \rho^2} dr^2) + (Q^2 d
\Omega^2).
\eeq
The geometry thus factorizes as for the Schwarzschild solution.
One factor the 2-sphere $S^2$ of radius $Q$ but the other  $(r,
t)$ factor is now not Rindler any more  but is  a two-dimensional
Anti-de Sitter or $AdS_2$. The geodesic radial distance in $AdS_2$
is $\log{r}$. As a result the geometry looks like an infinite
throat near $r =0$ and the radius of the  mouth of the throat has
radius $Q$.

Extremal black holes  are interesting because they are stable
against Hawking radiation and nevertheless have a large entropy.
We  now try to see if the entropy can be explained by counting of
microstates. In doing so, supersymmetry proves to be a very useful
tool.

\section{Wald entropy for extremal black holes \label{EntropyFunction}}

The horizon of extremal black holes has additional symmetries.  For non-spinning  black holes, the geometry is spherically symmetric.  At extremality,  the near horizon geometry becomes  $AdS_{2} \times S^{2}$ just as in the case of Reissner-Nordstr\"om black hole. The formula for the Wald entropy can be simplified considerably by exploiting these symmetries \cite{Sen:2008yk, Sen:2008vm}. 

 The Reissner-Nordstr\"om metric is 
\be \label{em1}
ds^2  = - (1 -  r_{+}/ r) (1 -  r_{-}/ r) dt^2   + {d r^2\over
 (1 -r_{+}/ r) (1 -  r_{-}/ r)}   
 +  r^2 (d\theta^2 + \sin^2\theta d\phi^2)\, . 
 \ee
Here $(t, r,\theta,\phi)$ are the coordinates of space-time and
$ r_{+}$ and $ r_{-}$ are two parameters labelling the positions
of the outer and inner horizon of the black hole respectively ($ r_{+}> r_{-}$).
The extremal limit corresponds to $ r_{-}\to  r_{+}$. 
We take this limit keeping the
coordinates $\theta$, $\phi$, and
  \be \label{em2}
\sigma := \frac{\left(  2r-  r_{+} - r_{-}\right)}{( r_{+}- r_{-})} , \quad
\t := \frac{( r_{+}- r_{-}) t}{2 r_{+}^2 }\, , 
\ee
fixed. 
In this limit the metric and the other fields take the form:
\be \label{et1}
ds^2 =  r_{+}^2\left(-(\s^2-1) d\t^2 + {d\s^2\over \s^2-1}\right)
+  r_{+}^2\left(d\theta^2 + \sin^2(\theta) d\phi^2 \right)\, . \ee
This is the metric of $AdS_2\times S^2$, with
$AdS_2$ parametrized by $(\s, \t)$ and $S^2$ parametrized by
$(\theta,\phi)$.
Although in the original coordinate system the
horizons coincide in the extremal limit, in the
$(\s, \t)$ coordinate system the two horizons are
at $\s=\pm 1$.
The $AdS_{2}$ space has $SO(2, 1) \equiv SL(2, \mathbb{R})$ symmetry-- the time translation symmetry is enhanced to the larger $SO(2, 1) $ symmetry. All known extremal black holes have this property. Henceforth,  we will take this as a definition of the near horizon geometry of an extremal black hole. In four dimensions, we also have the $S^{2}$ factor with $SO(3)$ isometries. 
Our objective will be to exploit the $SO(2, 1) \times SO(3)$ isometries of this spacetime to considerably simply the formula for Wald entropy. 

Consider an arbitrary theory of gravity in four spacetime dimensions with metric $g_{\mu\nu}$  coupled to a set of
$U(1)$ gauge fields $A^{(i)}_{\mu}$  ($ i= 1, \ldots, r$ for a rank $r$ gauge group) and neutral scalar fields $\phi_{s}$  ($s=1, \ldots N$) . Let  $x^{\mu}$ ($\mu = 0, \ldots, 3$ be local coordinates on spacetime and  $\mathcal{L}$ be an arbitrary general coordinate invariant local lagrangian. The action is then
   \begin{equation}
I = \int d^{4}x \sqrt{-det(g) } \, \mathcal{L} \, . 
\end{equation}
For an extremal black hole solution of this action,
the most general form of the near horizon geometry and of all other fields consistent
with $SO(2, 1) \times  SO(3)$ isometry
 is given by
\begin{eqnarray}
ds^2 &=&  v_{1}\left(-(\s^2-1) d\t^2 + {d\s^2\over \s^2-1}\right)
+  v_{2} (d\theta^2 + \sin^2(\theta )d\phi^2)\,  , \\
F^{(i)}_{\s\t} &=& e_{i} \,\, , \quad F^{(i)}_{\theta\phi} = \frac{p_{i}}{4\pi} \sin{(\theta)} \,\, , \quad \phi_{s} = u_{s} \, \,  .
\end{eqnarray}
We can think of $e_{i}$ and $p_{i}$ $(i = 1, \ldots, r)$ as the electric and magnetic fields respectively near the black hole horizon. The constants $v_{a}$ ($a =1, 2$) and $u_{s}$ ($s = 1, \ldots, N$) are to be determined by solving the equations of motion. Let us  define
\begin{equation}
f(u, v, e, p) := \int d\theta d\phi \sqrt{-\det(g)} \mathcal{L}|_{horizon} \, .
\end{equation}
Using the fact that $\sqrt{-\det(g)} = \sin(\theta)$ on the horizon,
we conclude
\begin{equation} \label{fhorizon}
f(u, v, e, p) :=4\pi v_{1} v_{2} \mathcal{L}|_{horizon}
\end{equation}
Finally we define the entropy function
\begin{equation}
\mathcal{E}(q, u, v, e, p) = 2\pi (e_{i }q_{i} - f(u, v, e, p) ) \, ,
\end{equation}
where we have introduced the quantities 
\begin{equation}\label{qdef} 
q_{i} :=  \frac{\partial f}{\partial e_{i}}
\end{equation}
 which by definition can be  identified with the electric charges carried by the black hole. This function called the `entropy function' is directly related to the Wald entropy as we summarize below.

\begin{enumerate}
\item For a black hole with fixed electric charges $ \{q_{i}\}$ and magnetic charges $\{p_{i}\}$,  all  near horizon parameters $v, u, e$ are determined  by extremizing $\mathcal{E}$ with respect
to the near horizon parameters:
\begin{eqnarray}
{\partial \mathcal{E}\over \partial e_{i}}&=& 0 \quad  i=1, \ldots r  \, ; \label{ex1}\\
{\partial \mathcal{E}\over \partial v_{a}} &=& 0, \quad a = 1, 2; \label{ex2} \\{\partial  \mathcal{E}\over \partial  u_{s}}&=& 0,  \quad s = 1, \ldots N\, . \label{ex3}
\end{eqnarray}
Equation \eqref{ex1} is simply the definition of electric charge whereas the other two equations \eqref{ex2} and \eqref{ex3} are the equations of motion for the near horizon fields. This follows from the fact that 
the dependence of $\mathcal{E}$ on all the near horizon parameters
other than $e_{i}$ comes only through $f(u, v, e, p)$ which from \eqref{fhorizon} is proportional to the action near the horizon. Thus extremization of the near horizon action is the same as the extremization of $\mathcal{E}$.
This determines the variables $(u, v, e)$ in terms of $(q, p)$ and as a result the value of the entropy function at the extremum $\mathcal{E}^{*}$ is a function only of the charges
\begin{equation}
\mathcal{E}^{*}(q, p) := \mathcal{E}(q, u^{*}(q, p), v^{*}(q, p), e^{*}(q, p), p) \, .
\end{equation}
\item Once we have determined the near horizon geometry, we can find the entropy using  Wald's formula specialized to the case of extermal black holes:
\begin{equation}
S_{wald} = -8\pi \int d\theta d\phi \frac{\partial{S}}{\partial R_{rtrt}} \sqrt{-g_{rr}g_{tt}} \, .
\end{equation}
With some algebra it is easy to see that the entropy is  given by the value of the entropy function at the extremum:
\be \label{et5}
S_{wald}( q, p) = \mathcal{E}^{*}(q, p)\, .
\ee
\end{enumerate}

 This  `entropy function formalism'  described above  allows one to compute the entropy of various extremal black holes very efficiently by simply solving certain algebraic equations (instead of partial differential equations). It also allows one to incorporate effects of higher derivative corrections to the two-derivative action with relative ease.

\subsubsection{Wald entropy for a Reissner-Nordstr\"om black hole}

To illustrate the use of the entropy function formalism for concrete computations, consider the Einstein-Maxell theory given by the action (\ref{action}) and a solution given by
\begin{eqnarray}
 ds^2&=&v_{1}\left(-(\sigma^2-1)d\t^2+\frac{d\sigma^2}{\sigma^2-1}\right)\,\, + v_{2}\left(d\theta^{2} + sin^{2}(\theta) d\phi^{2}\right)  \nonumber \\
&& F_{\sigma \t}= e \, , \quad F_{\theta\phi} =  \frac{p}{4\pi} \sin{(\theta})\end{eqnarray}
Substituting into the action we obtain the entropy function
\begin{eqnarray}\label{eag7pre}
\mathcal{E}( q, v,  e,
 q, p) &\equiv& 2\pi \left( e_i q_i 
- f( v,  e, p) \right) \nonumber \\
&=& 2\pi \Bigg[ e q  -  \, 4 \pi v_1 \, v_2
\,  \bigg\{{1\over 16\pi}\left( -{2\over v_1} +{2\over v_2} \right)+ 
 {1\over 2 v_1^2} e^{2}
- {1\over 32\pi^2 v_2^2}  p^{2} \bigg\} \Bigg]\, .
\end{eqnarray}
The extremization equations
\begin{equation}
\frac{\partial \mathcal{E}}{\partial e} = 0 \, ,\quad \frac{\partial\mathcal{E}}{\partial v_{1}}=0 \, , \quad \frac{\partial \mathcal{E}}{\partial v_{2}} =0
\end{equation}
can be easily solved to obtain
\begin{equation}
v_{1} = v_{2} = \frac{q^{2} + p^{2}}{4\pi} \, , \quad e = \frac{q}{4\pi}
\end{equation}
and 
\begin{equation}
S_{wald}(q, p) = \mathcal{E}^{*}(q, p) = \frac{q^{2} + p^{2}}{4} \, .
\end{equation}

\chapter{Elements of String Theory \label{Elements}}

\section{BPS states in $\mathcal{N} =4$ string compactifications}

Superstring theories are naturally formulated in ten-dimensional Lorentzian
spacetime $ \mathcal{M}_{10}$. A `compactification' to four-dimensions is obtained by
taking $\mathcal{M}_{10}$ to be a product manifold $ \mathbb{R}^{1, 3} \times X_6$ where
$X_6$ is a compact Calabi-Yau threefold and $\mathbb{R}^{1, 3}$ is the noncompact
Minkowski spacetime. We will focus in these lectures on a compactification of Type-II
superstring theory when $X_6$ is itself the product $X_6 = K3 \times T^2$. A
highly nontrivial and surprising result from the 90s is the statement that this
compactification is quantum equivalent or `dual' to a compactification of
heterotic string theory on $T^4 \times T^2$ where $T^4$ is a four-dimensional
torus \cite{Hull:1994ys, Witten:1995ex}. One can thus describe the theory either
in the Type-II frame or the heterotic frame.

The four-dimensional theory in $\mathbb{R}^{1, 3}$ resulting from this
compactification has $ \mathcal{N}=4$ supersymmetry\footnote{This supersymmetry is a super Lie algebra containing $ISO(1, 3) \times SU(4)$ as the bosonic subalgebra where $ISO(1,3)$ is the Poincar\'e symmetry of the $\mathbb{R}^{1, 3}$ spacetime and $SU(4)$ is an internal symmetry called R-symmetry in physics literature. The odd generators of the superalgebra are called supercharges. With $ \mathcal{N}=4$ supersymmetry, there are eight complex supercharges which transform as a spinor of $ISO(1,3)$ and a fundamental of $SU(4)$.}. The massless fields in the theory consist of $22$ vector multiplets in addition to the supergravity
multiplet. The massless moduli fields consist of the S-modulus $\lambda$ taking
values in the coset
\begin{equation}\label{Smoduli}
 SL(2, \mathbb{Z})\backslash SL(2; \mathbb{R})/ O(2; \mathbb{R}),
\end{equation}
and the T-moduli $\mu$ taking values in the coset
\begin{equation}\label{Narainmoduli}
   O(22, 6; \mathbb{Z}) \backslash O(22, 6; \mathbb{R}) /O(22; \mathbb{R}) \times
O(6; \mathbb{R}). \end{equation} The group of discrete identifications $SL(2,
\mathbb{Z})$ is called S-duality group. In the heterotic frame, it is the
electro-magnetic duality group \cite{Sen:1994yi, Sen:1994fa} whereas in the
type-II frame, it is simply the group of area- preserving global diffeomorphisms
of the $T^{2}$ factor. The group of discrete identifications $O(22, 6;
\mathbb{Z})$ is called the T-duality group. Part of the T-duality group $O(19, 3;
\mathbb{Z})$ can be recognized as the group of geometric identifications on the
moduli space of K3; the other elements are stringy in origin and have to do with
mirror symmetry.

At each point in the moduli space of the internal manifold $K3 \times T^2$, one
has a distinct four- dimensional theory. One would like to know the spectrum of
particle states in this theory. Particle states are unitary irreducible
representations, or supermultiplets, of the $\mathcal{N} =4$ superalgebra. The
supermultiplets are of three types which have different dimensions in the rest
frame. A long multiplet is $256$- dimensional, an intermediate multiplet is
$64$-dimensional, and a short multiplet is $16$- dimensional. A short multiplet
preserves half of the eight supersymmetries (\textit{i.e.} it is annihilated by four
supercharges) and is called a half-BPS state; an intermediate multiplet preserves
one quarter of the supersymmetry (\textit{i.e.} it is annihilated by two
supercharges), and is called a quarter-BPS state; and a long multiplet does not
preserve any supersymmetry and is called a non-BPS state. One consequence of the
BPS property is that the spectrum of these states is `topological' in that it
does not change as the moduli are varied, except for jumps at certain walls in
the moduli space \cite{Witten:1978mh}.

An important property of the BPS states that follows from the superalgebra is
that their mass is determined by the charges and the moduli \cite{Witten:1978mh}.
Thus, to specify a BPS state at a given point in the moduli space, it suffices to
specify its charges.  The charge vector in this theory transforms in the vector
representation of the T-duality group $O(22, 6; \mathbb{Z})$ and in the
fundamental representation of the S-duality group $SL(2, \mathbb{Z})$. It is thus
given by a vector $\G^{i\a}$  with integer entries
\begin{equation}\label{chargevector}
    \Gamma^{i\a} =
    			\left(
                             \begin{array}{c}
                               Q^i \\
                               P^i\\
                             \end{array}
                           \right) \quad where \quad \quad {{i=1,2, \ldots
28};\quad  \a = 1,2} \,
\, 
\end{equation}
transforming  in the $(2, 28)$ representation of $SL(2, \mathbb{Z}) \times O(22, 6; \mathbb{Z})$.
The vectors $Q$ and $P$ can be regarded as the quantized electric and magnetic
charge vectors of the state respectively. They both belong to an even, integral,
self-dual lattice $\Pi^{22, 6}$. We will assume in what follows that $ \Gamma =
(Q, P)$ in (\ref{chargevector}) is primitive in that it cannot be written as an
integer multiple of $ (Q_0, P_0)$ for $Q_0$ and $P_0$ belonging to $\Pi^{22, 6}$.
A state is called purely electric if only $Q$ is non-zero, purely magnetic if
only $P$ is non- zero, and dyonic if both $P$ and $Q$ are non-zero.

To define S-duality transformations, it is convenient to represent the S-modulus
as a complex field $S$ taking values in the upper half plane. An S-duality
transformation
\begin{equation}\label{Sgroup}
    \gamma \equiv\left(
      \begin{array}{cc}
        a&b\\
        c&d
      \end{array}
    \right) \in SL(2; \mathbb{Z})
\end{equation}
acts simultaneously on the charges and the S-modulus by
\begin{equation}\label{stransform}
    \left(
                             \begin{array}{c}
                               Q \\
                               P \\
                             \end{array}
                           \right) \rightarrow
    \left(
      \begin{array}{cc}
        a&b\\
        c&d
      \end{array}
    \right)
    \left(
                             \begin{array}{c}
                               Q \\
                               P \\
                             \end{array}
                           \right); \qquad S\to \frac{a S+b}{c S+d}
\end{equation}

To define T-duality transformations, it is convenient to represent the T-moduli
by a $28 \times 28$ of matrix $\mu^A_I$ satisfying \begin{equation} \mu^{t} \, L \,
\mu = L \end{equation} with the identification that $ \mu \sim k \mu  $ for
every $k\in O(22; \mathbb{R}) \times O(6; \mathbb{R})$. Here $L$ is the $(28
\times 28)$ matrix
\begin{equation}\label{lorentzian}
    L_{IJ} = \left(
               \begin{array}{ccc}
                 - \textbf{C}_{16} & \textbf{0} & \textbf{0}\\
                 \textbf{0} & \textbf{0} & \textbf{I}_{6}  \\
                 \textbf{0} &   \textbf{I}_{6} & \textbf{0}  \\
               \end{array}
             \right), \end{equation}
with $\textbf{I}_s$ the $s\times s$ identity matrix and $ \textbf{C}_{16}$ is the Cartan matrix of  $E_{8 }\times E_{8}$ . The T-moduli are then represented by the matrix 
\begin{equation}
\mathcal{M} =\mu^{t} \mu
\end{equation}
which satisifies
\begin{equation}\label{mconstraints}
\mathcal{M}^{t } = \mathcal{M}, \qquad \mathcal{M}^{t} L \mathcal{M} = L
\end{equation}
In this basis, a T-duality
transformation can then be represented by a $(28 \times 28)$ matrix $R$ with
integer entries satisfying
\begin{equation}\label{defRtduality}
    R^{t} L R = L ,
\end{equation}
which acts simultaneously on the charges and the T-moduli by
\begin{equation}\label{ttransform}
    Q \rightarrow R Q; \quad P \rightarrow R P ; \quad \mu \rightarrow   \mu R^{-1}
\end{equation}

Given the matrix $\mu^{A}_I$, one obtains an embedding $ \Lambda^{22, 6}\subset
\mathbb{R}^{22, 6}$ of $\Pi^{22, 6}$ which allows us to define the
moduli-dependent charge vectors $Q$ and $P$ by
\begin{equation}\label{physcharges}
    Q^{A} =  \mu^{A}_{I}Q_{I} \,   \quad P^{A} =  \mu^{A}_{I}P_{I} \, .
\end{equation}
Note that while $Q^{I}$ are integers $Q^{A}$ are not. In what follows we will not always write the indices explicitly assuming that it will be clear from the context. In any case, the final answers will only depend on the T-duality invariants which are all integers. 
The matrix $L$ has a $22$-dimensional eigensubspace with eigenvalue $-1$ and a
$6$- dimensional eigensubspace with eigenvalue $ +1$. Given $Q$ and $P$, one can
define the `right-moving' charges\footnote{The right- moving charges couple to
the graviphoton vector fields associated with the right-moving chiral currents in
the conformal field theory of the dual heterotic string.} $Q_R$ and $P_R$ as the
projections of $Q$ and $P$ respectively onto the subspace with eigenvalue $+1\,$. and the `left-moving' charges as  projections onto the subspace with eignevalue $-1\,$. These definitions can be compactly written as 
\begin{equation}
Q_{R, L} =\frac{(1 \pm L)}{2}  Q \, ; \quad P_{R, L} = \frac{(1 \pm L)}{2} P
\end{equation}
The right-moving charges since for the heterotic string,  $Q_{R}$  are related to the right-moving momenta. The central charges $Z_{1}$ and $Z_{2}$  of the ${\cal N}\,=\,4$  superalgebra can then be defined in terms of the right-moving charges and moduli (For details of these definitions and the superalgebra, see  \S\ref{BPS} ).

If the vectors $Q$ and $P$ are nonparallel, then the state is quarter-BPS. On the
other hand, if $Q= p Q_0$ and $ P =q Q_0$ for some $Q_0 \in \Pi^{22, 6}$ with $p$
and $q$ relatively prime integers, then the state is half-BPS.

An important piece of nonperturbative information about the dynamics of the
theory is the exact spectrum of all possible dyonic BPS- states at all points in
the moduli space. More specifically, one would like to compute the number
$d(\Gamma)|_{\l, \m}$ of dyons of a given charge $\Gamma$ at a specific point
$(\l, \m)$ in the moduli space. Computation of these numbers is of course a very
complicated dynamical problem. In fact, for a string compactification on a
general Calabi-Yau threefold, the answer is not known. One main reason for
focusing on this particular compactification on $K3 \times T^2$ is that in this
case the dynamical problem has been essentially solved and the exact spectrum of
dyons is now known. Furthermore, the results are easy to summarize and the
numbers $d(\Gamma)|_{\l, \m}$ are given in terms of Fourier coefficients of
various modular forms.

In view of the duality symmetries, it is useful to classify the inequivalent
duality orbits labeled by various duality invariants. This leads to an
interesting problem in number theory of classification of inequivalent duality
orbits of various duality groups such as $SL(2, \mathbb{Z}) \times O(22, 6; \mathbb{Z})$
in our case and more exotic groups like $E_{7,7} (\mathbb{Z})$ for other choices of
compactification manifold $X_6$. It is important to remember though that a
duality transformation acts simultaneously on charges and the moduli. Thus, it
maps a state with charge $\Gamma$ at a point in the moduli space $ (\l, \mu)$ to
a state with charge $ \Gamma'$ but at some other point in the moduli space $(\l',
\mu')$.  In this respect, the half-BPS and quarter-BPS dyons behave differently.
\begin{itemize}
  \item For half-BPS states, the spectrum does not depend on the moduli. Hence
$d(\Gamma)|_{\l', \m'} = d(\Gamma)|_{\l, \m}$. Furthermore, by an S-duality
transformation one can choose a frame where the charges are purely electric with
$P=0$ and $Q \neq 0$. Single-particle states have $Q$ primitive and the number of
states depends only on the T-duality invariant integer $n \equiv Q^2/2$. We can
thus denote the degeneracy of half-BPS states $d(\Gamma)|_{S', \m'}$ simply by
$d(n)$.
  \item For quarter-BPS states, the spectrum does depend on the moduli, and
$d(\Gamma)|_{\l', \m'} \neq d(\Gamma)|_{\l, \m}$. However, the partition function
turns out to be independent of moduli and hence it is enough to classify the
inequivalent duality orbits to label the partition functions. For the specific
duality group $SL(2, \mathbb{Z}) \times O(22, 6; \mathbb{Z})$ the partition functions are
essentially labeled by a single discrete invariant
\cite{Dabholkar:2007vk, Banerjee:2007sr, Banerjee:2008ri}.
      \begin{equation}\label{gcd}  I = \gcd (Q \wedge P) \, ,
      \end{equation}
The degeneracies themselves are Fourier coefficients of the partition
function. For a given value of $I$, they depend only on\footnote{There is an additional dependence on  arithmetic T-duality invariants but the degeneracies for states with nontrivial values of these T-duality invariants can be obtained from the degeneracies discussed here by demanding S-duality invariance \cite{Banerjee:2008ri}. } the moduli and the three
T-duality invariants $(m, n, \ell) \equiv (P^2/2, Q^2/2, Q \cdot P)$. Integrality
of $ (m, n, \ell)$ follows from the fact that both $Q$ and $P$ belong to
$\Pi^{22, 6}$. We can thus denote the degeneracy of these quarter-BPS states
$d(\Gamma)|_{\l, \m}$ simply by $d(m, n, l)|_{\l, \m}$.  For simplicity, we consider only $I=1$ in these lectures. Generalization for higher $I$ can be found in \cite{Banerjee:2008pu, Dabholkar:2008zy}. 
\end{itemize}

\section{Exercises}

\subsubsection{Elements of string compactifications}

The heterotic string theory in ten dimensions has $16$ supersymmetries. The bosonic massless fields consist of the metric $g_{MN}$, a 2-form field $B^{(2)}$,   $16$ abelian 1-form gauge fields $A^{(r)}$ $r=1, \ldots 16$, and a real scalar field $\phi$ called the dilaton. The Type-IIB string theory in ten dimensions has $32$ supersymmetries. The bosonic massless fields consist of the metric $g_{MN}$; two  2-form fields $C^{(2)}, B^{(2)}$;  a self-dual 4-form field $C^{(4)}$; and a complex scalar field $\lambda$ called the dilaton-axion field.

One of the remarkable strong-weak coupling dualities is the `string-string' duality between heterotic string compactified on $T^{4}\times T^{2}$ and Type-IIB string compactified on $K3 \times T^{2}$. One piece of evidence for this duality is obtained by comparing the massless spectrum for these compactifications and certain half-BPS states in the spectrum.
\begin{myenumerate}
\item 
\textit{Show that the heterotic string compactified on $T^{4}\times S^{1} \times \tilde S^{1}$ leads a four dimensional theory with $\mathcal{N} =4$ supersymmetry with $22$ vector multiplets.}
\item 
\textit{Show that the Type-IIB string compactified on $K3 \times S^{1} \times \tilde S^{1}$ leads a four dimensional theory with $\mathcal{N} =4$ supersymmetry with $22$ vector multiplets.}
\item \textit{Show that the Kaluza-Klein monopole in Type-IIB string associated with the circle $\tilde S^{1}$ has the right structure of massless fluctuations to be identified with the half-BPS perturbative heterotic string in the dual description.}
\end{myenumerate}

\section{String-String duality}

It will be useful to recall a few details of the string-string duality between heterotic compactified on $T^{4} \times S^{1} \times \tilde S^{1}$ and Type-IIB compactified on $K3 \times S^{1}\times \tilde S^{1}$. 
Two pieces of evidence for this duality will be relevant to our discussion.

$\bullet$ \textit{Low energy effective action}

Both these compactifications result in $\mathcal{N}=4$ supergravity in four dimensions. With this supersymmetry, the two-derivative effective action for the massless fields receives no quantum corrections. Hence, if the two theories are to be dual to each other, they must have identical 2-derivative action. 

This is indeed true. Even though the field content and the action are very different for the two theories in ten spacetime dimensions, upon respective compactifications, one obtains $\mathcal{N}=4$ supergravity 
 with  $22$ vector multiplets coupled to the supergravity multiplet. This has been discussed briefly in one of the tutorials.  For a given number of vector multiplets, the two-derivative action is then completely fixed by supersymmetry and hence is the same for the two theories. This was one of the properties that led to the conjecture of a strong-weak coupling duality between the two theories.  

For our purposes, we will be interested in the  2-derivative action for the bosonic fields. This   is a generalization of  the Einstein-Hilbert-Maxwell action \eqref{action} which couples the metric, the moduli fields and $28$ abelian gauge fields:
\begin{eqnarray}\label{hetaction}
 I&=&\frac{1}{32\pi}\int d^4x \sqrt{-\text{det}G}\, S \, [R_G +\frac{1}{S^2}G^{\mu\nu}(\partial_{\mu}S\partial_{\nu}S -\frac{1}{2}\partial_{\mu}a\partial_{\nu}a) \nonumber\\
&& \qquad\qquad + \frac{1}{8}G^{\mu\nu}Tr(\partial_{\mu}ML\partial_{\nu}ML) -G^{\mu\mu'}G^{\nu\nu'}F^{(i)}_{\mu\nu}(LML)_{ij}F^{(j)}_{\mu'\nu'}  \\
&&-\frac{a}{S}G^{\mu\mu'}G^{\nu\nu'}F^{(i)}_{\mu\nu}L_{ij}\tilde{F}^{(j)}_{\mu'\nu'}] \quad \quad  i, j = 1, \ldots, 28. \nonumber
\end{eqnarray}
In the heterotic string picture, the expectation value of the dilaton field $S$ is related to the four-dimensional string coupling $g_{4}$
\be \label{Smoduli1}
S \sim \frac{1}{g_{4}^{2}} \, ,
\ee
and $a$ is the axion field. 
The metric $G_{\m\nu}$ is the metric in the string frame and is related to the metric $g_{\mu\nu}$ in Einstein frame by the Weyl rescaling
\begin{equation}
g_{\m\nu} = S G_{\m\nu}
\end{equation}

$\bullet$ \textit{BPS spectrum}

Another requirement of duality is that the spectrum of BPS states should match for the two dual theories. Perturbative states in one description will generically get mapped to some non-perturbative states in the dual description. As a result, this leads to highly nontrivial predictions about the nonpertubative  spectrum in the dual description given the perturbative spectrum in one description.

As an example, consider the perturbative BPS-states in heterotic string theory on $K3 \times S^1 \times \tilde{S^1}$.  A heterotic string wrapping $w$ times on $S^{1}$ and carrying momentum $n$ gets mapped in Type-IIA to the NS5-brane wrapping  $w$ times on $K3 \times S^{1}$ and carrying momentum $n$. One can go from Type-IIA to Type-IIB by a T-duality along the $\tilde S^{1}$ circle. Under this T-duality, the  NS5-brane gets mapped to a KK-monopole with monopole charge   $w$ associated with the circle $\tilde S^{1}$ and carrying momentum $n$. This thus leads to a prediction that the spectrum of KK-monopole carrying momentum in Type-IIB should be the same as the spectrum of perturbative heterotic string discussed earlier. We will verify this highly nontrivial prediction in the next subsection for the case of $w=1$. 

\section{Kaluza-Klein monopole and the heterotic string}

The metric of the Kaluza-Klein monopole is given by the so called Taub-NUT metric
\be\label{tnutgeom}
ds_{TN}^2 = \left(1+\frac{R_0}{r}\right) 
\left( dr^2 +  r^2 ( d\theta^2 + \sin^2 \theta d\phi^2) \right)
+ R_0^2\left( 1 + \frac{R_0}{r}
\right)^{-1} ( 2\, d\psi + \cos\theta d\phi)^2
\ee
with the identifications:
\be\label{eident}
(\theta,\phi,\psi) \equiv (2\pi -\theta,\phi+\pi, \psi+{\pi\over 2})
\equiv (\theta,\phi+2\pi,\psi+\pi)\equiv (\theta,\phi,\psi+2\pi)\, .
\ee
Here $R_0$ is a constant determining the size of the Taub-NUT space $\mathcal{M}_{TN}$. This metric satisfies the Einstein equations in four-dimensional Euclidean space. The metric  \eqref{tnutgeom}
admits a normalizable self-dual harmonic form $\omega$, given
by
\be \label{nneomegadefrp}
\omega^{KK} = 
{r\over r +R_0}
d\sigma_3 + {R_0\over (r+R_0)^2}dr
\wedge \sigma_3\, , \qquad \sigma_3 \equiv
\left (d\psi + {1\over 2} \cos\theta d\phi\right)\, .
\ee

We are interested in the Type-IIB string theory
compactified on 
$K_{3}\times \tilde S^1\times  S^1$ in the presence of
a Kaluza-Klein monopole, with $\tilde S^1$ identified with the
asymptotic circle of the Taub-NUT space labeled by the coordinate
$\psi$ in \eqref{tnutgeom}. Thus, we want analyze the massless fluctuations of Type-IIB string on $K_{3}\times   S^1 \times \mathcal{M}_{TN} $ space.   Let $y$ and $\tilde y$ be the coordinates of $S^{1}$ and $\tilde S^{1}$ respectively with $y \sim y + 2\pi R$ and $\tilde y \sim \tilde y +2 \pi \tilde R$.
When the radius $R$  of the $S^{1}$ is large compared to the size of the $K3$ and the radius $\tilde R$ of the $\tilde S^{1}$ circle, we obtain an  `effective string' wrapping the $S^{1}$ with massless spectrum that agrees with the massless spectrum of a fundmental heterotic string wrapping $S^{1}$. These massless modes can be deduced as follows:
\begin{myitemize}
\item 
The center-of-mass of the KK-monopole can be located anywhere in $\mathbb{R}^{3}$ and its position is specified by a vector $\vec{a}$. Thus, we have
\begin{equation}
r := |\vec{x} - \vec{a}|  \, , \quad  \cos{\theta} := \frac{x^{3}-a^{3}}{r} \, , \quad \tan{\phi} := \frac{x^{1}-a^{1}}{x^{2} -a^{2}} \, .
\end{equation}
if $(x^{1}, x^{2}, x^{3})$ are the coordinates of $\mathbb{R}^{3}$. 
We can allow these coordinates to fluctuate in the $t$ and $y$ directions and hence we will obtain
three non-chiral massless $a^{i}(t, y)$
scalar fields  along the effective string associated with oscillations
of the three coordinates of the center-of-mass of the KK monopole.
\item
There are 
two additional non-chiral scalar 
fields $b(t, y)$ and $c(t, y)$ obtained by reducing the two 2-form fields $B^{(2)}$ and $C^{2}$  of Type-IIB
along
the harmonic
2-form \eqref{nneomegadefrp}:
\begin{equation}
B^{(2)} = b(t, y) \cdot \omega^{KK} \, \quad C^{(2)} =  c(t, y) \cdot \omega^{KK}
\end{equation}
\item 
There are 
3 right-moving $a_{R}^{r}(t +y) \, , r =1, 2, 3$ and $19$ left-moving scalars $a_{L}^{s}(t-y) \, , s =1, \ldots, 19$ obtained by reducing the self-dual 4-form field $C^{(4)}$ of type IIB theory. This works as follows. The  field $C^{(4)}$ can be reduced taking it as a tensor product of the harmonic
2-form \eqref{nneomegadefrp} and a harmonic 2-form $\omega^{K_{3}}_{\alpha}$ for $\alpha = 1, \ldots, 22$ on $K_{3}$. This gives rise to
rise to a chiral scalar field on the world-volume. The chirality of
the scalar field is correlated with whether the corresponding
harmonic 2-form $\omega^{K_{3}}_{\alpha}$ is self-dual or anti-self-dual.
Since  $K3$ has three self-dual $\omega^{K_{3}+}_{r}$ and nineteen 
anti-selfdual harmonic 2-forms $\omega^{K_{3}-}_{s}$, we get $3$ right-moving and $19$ left-moving scalars:
\begin{equation}
C^{(4)} = \sum_{r=1}^{3 }a_{R}^{s}(t + y) \cdot \omega^{K_{3}-}_{s}\wedge \omega^{KK} + \sum_{s=1}^{19 }a_{L}^{s}(t-y)\cdot \omega^{K_{3}-}_{s}\wedge \omega^{KK} \, .
\end{equation}
\end{myitemize}
The KK-monopole background breaks $8$ of the $16$ supersymmetries of Type-II on $K3 \times S^{1}$. Consequently,  there are eight right-moving fermionic fields
$$S^{a}(t+y) \quad a= 1, \ldots, 8 $$
which arise as the goldstinos of these eight broken supersymmetries. This is precisely the field content of the $1+1$ dimensional worldsheet theory of the heterotic string wrapping $S^{1}$ as we discussed in the tutorial \eqref{countingsec}.

\section{Supersymmetry and extremality}

Some of the special properties of external black holes  can be
understood better by embedding them in  supergravity. We will be interested in these lectures in string compactifications with  $\mathcal{N} =4$ supersymmetry in four spacetime dimensions.
The $\mathcal{N} =4$ supersymmetry algebra contains in addition to the usual
Poincar\'e generators, sixteen real supercharges which can be grouped into 8 complex charges  $Q_\alpha^a$ and their complex conjugates. Here $\alpha
= 1, 2$ is the usual Weyl spinor index of 4d Lorentz symmetry.
and the  internal index
$a=1, \ldots, 4$  in the fundamental $\textbf{4}$ representation of an
$SU(4)$, the R-symmetry of the superalgebra. The relevant
anticommutators for our purpose are
\begin{eqnarray} 
\{Q_\alpha^a, {\bar Q_{\dot{\beta} b}}\} &=&-
2 P_\mu \sigma^\mu_{\alpha\dot{\beta}}\delta^a_b \label{one1} \nonumber \\
\{Q_\alpha^a, {Q_{\beta}^b}\} =  \epsilon_{\alpha\beta}
Z^{ab}
 &\quad&
 \{{\bar Q_{\dot{\alpha}a}}, {\bar Q_{\dot{\beta }b}}\}
 ={\bar Z}_{ab} \epsilon_{{\dot \alpha}{\dot \beta}}
 \label{two2}
\end{eqnarray}
where $\sigma^\mu$ are $( 2\times 2)$ matrices with $\sigma_0 =
-\textbf{1}$ and $\sigma^i for i= 1,2, 3$ are the usual Pauli matrices.
Here $P_\mu$ is the momentum operator and $Q$ are the
supersymmetry generators and the complex number $Z^{ab}$ is the central
charge matrix. 

Let us first look at the representations of this algebra when the
central charge is zero. In this case the massive and massless
representation are qualitatively different.

\begin{enumerate}
    \item Massive Representation, $M > 0, P^\mu = (M, 0, 0, 0)$\\
In this case,  (\ref{one1}) becomes $ \{Q_\alpha^a, {\bar
Q_{\dot{\beta}b}}\} = 2 M \delta_{\alpha\dot{\beta}}\delta^a_b$
and all other anti-commutators vanish. Up to overall scaling,
these are the commutation relations for eight complex fermionic
oscillators. Each oscillator has a two-state representation, which is either filled or empty. These states together define a unitary irreducible representation, called a supermultiplet,  of the superalgebra. 
The total dimension of the
representation is $2^8=256$ which is CPT self-conjugate.

\item Massless Representation $M =0, P^\mu = (E, 0, 0,
E)$\\
In this case (\ref{one1}) becomes $ \{Q_1^a, {\bar
Q_{\dot{1}b}}\} = 2 E \delta^a_b$ and all
other anti-commutators vanish.  Up to overall scaling, these are
now the anti-commutation relations of \emph{four} fermionic
oscillators and hence the total dimension of the representation is
$2^4 =16$ which is also CPT-self-conjugate. 
\end{enumerate}
The important point is that for a massive representation, with $M
=\epsilon >0 $, no matter how small $\epsilon$, the supermultiplet
is long and precisely at $M=0$ it is short. Thus the size of the
supermultiplet has to change discontinuously if the state has to
acquire mass.  Furthermore, the size of the supermultiplet is
determined by the number of supersymmetries that are \emph{broken}
because those have non-vanishing anti-commutations and turn into
fermionic oscillators.

Note that there is a
bound on the mass $M \geq 0$ which simply follows from the fact
the using (\ref{one1}) one can show that the mass operator on the
right hand side of the equation equals a positive operator, the
absolute value square of the supercharge on the left hand side.
The massless representation saturates this bound and is `small'
whereas the massive representation is long. 

There is an analog of
this phenomenon also for nonzero $Z_{ab}$.   
As explained in the appendix, the central charge matrix $Z_{ab}$ can be brought to the standard form by an $U(4)$ rotation
\begin{equation}
\tilde{Z} = U Z U^T, \;\; U \in U(4)\;, \;\;\;\;\; \tilde{Z}_{ab} = \left(\begin{array}{c|c} Z_1 \varepsilon & 0  \\   \hline  0 & Z_2 \varepsilon \end{array}\right)\;, \;\;\; \varepsilon = \left(\begin{array}{cc} 0 & 1  \\  -1 & 0 \end{array}\right) \label{diag} \, .
\end{equation}
so we have two `central charges' $Z_{1}$ and $Z_{2}$.
Without loss of generality we can assume $|Z_1| \geq |Z_2|$.
Using the supersymmetry algebra one can prove the
BPS bound $M - |Z_{1}| \geq 0 $  by showing that this operator is equal
to a positive operator (see appendix for details). States that saturate this bound are the BPS states.
 There are three types of representations:
\begin{itemize}
\item If $M= |Z_1|=|Z_2|$, then eight of of the sixteen supersymmetries are preserved. Such states are called half-BPS. The broken supersymmetries result in four complex fermionic zero modes whose quantization furnishes a $2^{4}$-dimensional short multiplet
\item 
If $M = |Z_1| > |Z_2|$, then and four out of the sixteen supersymmetries are preserved. Such states are called quarter-BPS. The broken supersymmetries result in six complex fermionic zero modes whose quantization furnishes a $2^{6}$-dimensional intermediate multiplet. 
\item
If $M > |Z_1| > |Z_2|$, then no supersymmetries are preserved. Such states are called non-BPS.The sixteen broken supersymmetries result in eight complex fermionic zero modes  whose quantization furnishes a $2^{8}$-dimensional  long multiplet.
\end{itemize}

The significance of BPS states in string theory and in gauge
theory stems from the classic argument of Witten and Olive which
shows that under suitable conditions, the spectrum of BPS states
is stable under smooth changes of moduli and coupling constants.
The crux of the argument is that with sufficient supersymmetry,
for example $\mathcal{N} = 4$, the coupling constant does not get
renormalized. The central charges $Z_{1}$ and $Z_{2}$ of the supersymmetry algebra
depend on the quantized charges and the coupling constant which
therefore also does not get renormalized. This shows that for BPS
states, the mass also cannot get renormalized because if the
quantum corrections increase the mass, the states will have to
belong a long representation . Then, the number of states will have
to jump discontinuously from, say from $16$ to $256$ which cannot
happen under smooth variations of couplings unless there is some kind of a `Higgs Mechanism' or there is some kind of a 
phase transition\footnote{Such `phase transitions' do occur and the degeneracies can jump upon crossing certain walls in the moduli space. This phenomenon called   `wall-crossing' occurs not because of Higgs mechanism but because at the walls,  single particle states have the same mass as certain multi-particle states and can thus mix with the multi-particle continuum states. The wall-crossing phenomenon complicates the analytic continuation of the degeneracy from weak coupling from strong coupling since one may encounter various walls along the way. However, in many cases, the jumps across these walls can be taken into account systematically.}

As a result, one can compute the spectrum at weak coupling in the
region of moduli space where perturbative or semiclassical
counting methods are available. One can then analytically continue
this spectrum to strong coupling. This allows us to obtain
invaluable non-perturbative information about the theory from
essentially perturbative commutations.

\section{BPS dyons in  $\mathcal{N}=4$ compactifications}

The massless spectrum  of the toroidally compactified heterotic string on $T^{6}$ contains $28$ different ``photons''  or $U(1)$ gauge fields -- one from each of the $22$ vector multiplets and $6$ from the supergravity multiplet. As a result, the electric charge of a state is specified by a $28$-dimensional charge vector $Q$ and the magnetic charge is specified by a $28$-dimensional charge vector $P$. Thus, a  dyonic state is specified by the charge vector
\begin{equation}\label{chargevec}
    \Gamma = \left(
               \begin{array}{c}
                 Q \\
                 P \\
               \end{array}
             \right)
\end{equation}
where $Q$ and $P$  are the electric and magnetic charge vectors respectively. Both $Q$ and $P$ are elements of a self-dual integral lattice $\Pi^{22, 6}$ and can be represented as $28$-dimensional column vectors  in $\mathbb{R}^{22, 6}$ with integer entries, which transform in the fundamental representation of $O(22, 6; \mathbb{Z})$.  We will be interested in BPS states. 
\begin{itemize}
\item
For  half-BPS state the charge vectors $Q$ and $P$ must be parallel. 
These states are dual to perturbative BPS states.
\item
For a quarter-BPS states the charge vectors $Q$ and $P$ are not parallel.
There is no duality frame in which these states are perturbative.
\end{itemize}
There are three  invariants of $O(22, 6; \mathbb{Z})$, quadratic in charges, and given by $P^{2}$,  $Q^{2}$ and $Q \cdot P$. These three T-duality invariants will be useful in later discussions.

\chapter{Spectrum of Half-BPS Dyons \label{Half}}

An instructive example of BPS of states is provided by an infinite
tower of BPS states that exists in perturbative string theory \cite{Dabholkar:1989jt, Dabholkar:1990yf}.

 \section{Perturbative half-BPS States} \label{countingsec}

Consider a perturbative heterotic string state wrapping around $S^1$ with
winding number $w$ and quantized momentum $n$. Let the radius of
the circle be $R$ and $\alpha' =1$, then one can define
left-moving and right-moving momenta as usual,
\begin{equation}\label{momenta}
    p_{L,R} = \sqrt{1 \over 2} \left({n \over R} \pm wR\right).
\end{equation}

Recall that the heterotic strings consists of a right-moving
superstring and a left-moving bosonic string. In the NSR formalism
in the light-cone gauge, the worldsheet fields are:
\begin{itemize}
    \item Right moving superstring $
X^i (\sigma^-) \ \ \tilde\psi^i (\sigma^-) \qquad i = 1 \cdots 8
$
\item Left-moving bosonic string $
X^i (\sigma^+), X^I (\sigma^+) \qquad I = 1 \cdots 16$, 
\end{itemize}
where $X^i$ are the bosonic transverse spatial coordinates,
$\tilde \psi^i$ are the worldsheet fermions, and  $X^I$ are the
coordinates of an internal $E_8 \times E_8$ torus. 
A BPS state is obtained by keeping the right-movers in the ground
state ( that is,  setting the right-moving oscillator number $\tilde N= \frac{1}{2}$ in the NS sector and $\tilde N =0$ in the R sector). 

The Virasoro constraints are then given by

\begin{eqnarray}\label{vir}
 \tilde{L_{0}} - {M^2 \over 4} +  {p^2_R \over 2} &=& 0 \\
L_{0}  - {M^2 \over 4} +  {p^2_L \over 2} &=& 0,
\end{eqnarray}
where $N$ and $\tilde N$ are the left-moving and right-moving
oscillation numbers respectively.

The left-moving oscillator number is then
\begin{equation}\label{left}
L_{0} = \sum_{n=1}^\infty (\sum_{i=1}^{8} n a^i_{-n} a^i_n +
\sum_{I=1}^{16} n \beta^I_{-n} \beta^I_{-n}) -1 := N -1,
\end{equation}
where $a^i$ are the left-moving Fourier modes of the fields
$X^i$, and $\beta^I$ are the Fourier modes of the fields $X^I$.
Note that the right-moving fermions satisfy anti-periodic
boundary condition in the NS sector and have half-integral moding,
and satisfy periodic boundary conditions in the R sector and have
integral moding. The oscillator number operator is then given by
\begin{equation}\label{ns}
     \tilde L_{0} = \sum_{n =1}^\infty \sum_{i=1}^8 ( n
\tilde a^i_{-n} \tilde a^i_n +  r \tilde \psi^i_{-r}
\tilde\psi^i_r - {1\over 2}) := \tilde N -\frac{1}{2}. 
\end{equation}
with $r \equiv {-(n -{1\over 2})}$ in the NS sector and by
\begin{equation}\label{ramond}
      \tilde L_{0} = \sum_{n =1}^\infty \sum_{i=1}^8 ( n
\tilde a^i_{-n} \tilde a^i_n +  r \tilde \psi^i_{-r}
\tilde\psi^i_r )
\end{equation}
with $r \equiv (n-1)$ in the R sector.

In the NS-sector then one then has $\tilde N = {1\over 2}$ and the
states are given by
\begin{equation}\label{nsground}
\tilde \psi_{-{1\over 2}}^i |0>,
\end{equation}
that transform as the vector representation $8_v$ of $SO(8)$.
In the R sector the ground state is furnished by the
representation of fermionic zero mode algebra $\{ \psi^i_0,
\psi^j_0\} = \delta^{ij}$ which after GSO projection transforms as
$8_s$ of $SO(8)$. Altogether the right-moving ground state is
thus 16-dimensional $8_v \oplus {8_s}$.
From the
Virasoro constraint (\ref{vir}) we see that a BPS state with $\tilde N =0$ saturates
the BPS bound
\begin{equation}\label{bps}
    M = \sqrt{2} p_R,
\end{equation}
and thus $\sqrt{2} p_R$ can be identified with the central charge
of the supersymmetry algebra. The right-moving ground state after
the usual GSO projection is indeed 16-dimensional as expected for
a BPS-state in a theory with $\mathcal{N}=4$ supersymmetry.

We thus have a perturbative BPS state  which looks pointlike in
four dimensions with two integral charges $n$ and $w$ that couple
to two gauge fields $g_{5\mu}$ and $B_{5\mu}$ respectively. It
saturates a BPS bound $M = \sqrt{2} p_R$ and belongs to a
16-dimensional short representation. This  point-like state is our
`would-be' black hole. Because it has a large mass, as we increase
the string coupling it would begin to gravitate and eventually
collapse to form a black hole.

Microscopically, there is a huge multiplicity of such states which
arises from the fact that even though the right-movers are in the
ground state, the string can carry arbitrary left-moving
oscillations subject to the Virasoro constraint. Using $M=
\sqrt{2}p_R$ in the Virasoro constraint for the left-movers gives
us
\begin{equation}\label{vir2}
    N -1=  {1 \over 2}(p_R^2 - p_L^2 ) := Q^{2}/2 = nw.
\end{equation}
We would like to know the degeneracy of states for a given value
of charges $n$ and $w$ which is given by exciting arbitrary
left-moving oscillations whose total worldsheet oscillator excitation number adds up to
$N$. Let us take $w=1$ for simplicity and denote the degeneracy by $d (n)$ which we want to
compute. As usual, it is more convenient to evaluate the canonical
partition function
\begin{eqnarray}
  Z(\beta) &=& \Tr \left( e^{ - \beta L_{0}} \right) \\
    &\equiv& \sum_{-1}^{\infty} d(n)q^{n} \quad q := e^{-\beta}\, .
\end{eqnarray}
This is the canonical partition function of $24$ left-moving massless bosons in $1+1$ dimensions at temperature $1/\beta$. The micro-canonical degeneracy $d(N)$ is given then given as usual by the inverse Laplace
transform
\begin{equation}\label{omega1}
    d (N) = {1\over 2\pi i} \int d\beta e^{\beta N} Z(\beta).
\end{equation}

 Using the expression (\ref{left}) for the
oscillator number $s$ and the fact that
\begin{equation}\label{single}
    \textrm{Tr}(q^{-s\alpha_{-n} \alpha_n}) = 1 + q^s + q^{2s} +
q^{3s} + \cdots = {1\over (1 - q^s)}\, ,
\end{equation}
the partition function can be readily evaluated to obtain
\begin{equation}\label{partition}
     Z(\beta) =  {1 \over q} \prod_{s=1}^\infty {1\over(1 -
  q^s)^{24}} \quad .
\end{equation}
It is convenient to introduce a variable $\tau$ by $ \beta := -2 \pi i \tau$ , so that$q:=e^{2\pi i \tau} $. The function 
\begin{equation}\label{partition2}
     \Delta(\tau) =  q \prod_{s=1}^\infty {(1 -
  q^s)^{24}}, \quad 
\end{equation}
is the  famous discriminant function. Under modular transformations
\begin{equation}
\tau \rightarrow  \frac{a\tau + b}{c\tau +d} \quad a, b, c, d \in \mathbb{Z} \, , \quad {with} \quad ad -bc = 1
\end{equation}
it transforms as  a modular form of weight $12$:
\begin{equation}
\Delta(\frac{a\tau + b}{c\tau +d} ) = (c\tau + d)^{12} \Delta (\tau) \, , 
\end{equation}
This remarkable property allows us to relate high temperature $(\beta \rightarrow 0)$  to low tempreature $(\beta \rightarrow \infty)$ and derive a simple explicit expression for the asymptotic degeneracies $d(n)$ for $n$ very large.

\section{Cardy formula}

The degeneracy $d(N)$ can be obtained from the canonical partition function by the inverse Laplace transform
\begin{equation}\label{omega}
    d (N) = {1\over 2\pi i} \int d\beta e^{\beta N} Z(\beta).
\end{equation}
We would like to evaluate this integral (\ref{omega}) for large $N$ which
corresponds to large worldsheet energy.  Such  an asymptotic expansion of $d(N)$ for large $N$ is given by the `Cardy formula' which utilizes the modular properties of the partition function.

For large $N$, we  expect
that the integral receives most of its contributions from high
temperature or small $\beta$ region of the integrand.  To compute
the large $N$ asymptotics,  we then need to  know the small
$\beta$ asymptotics of the partition function. Now, $\beta
\rightarrow 0$ corresponds to $q \rightarrow 1$ and in this limit
the asymptotics of $Z(\beta)$ are very difficult to read off from
(\ref{partition}) because its a product of many quantities that
are becoming very large. It is more convenient to use the fact
that $Z(\beta)$ is the inverse of $\Delta(\tau)$ which is a modular form of weight $12$  we can conclude
\begin{equation}\label{modprop}
    Z(\beta)  = (\beta/2\pi)^{12} Z ({4\pi^2 \over \beta}).
\end{equation}
This allows us to relate the $q \rightarrow 1$ or high temperature
asymptotics to $q \rightarrow 0$ or low temperature asymptotics as
follows. Now, $Z(\tilde\beta) = Z \left({4\pi^2 \over
\beta}\right)$ asymptotics are easy to read off because as $\beta
\to 0$ we have $\tilde\beta \to \infty$ or $e^{-\tilde\beta} =
\tilde q \to 0$. As $\tilde q \to 0$
\begin{equation}\label{asympto}
    Z(\tilde\beta) = {1\over \tilde q}\prod_{n=1}^\infty {1\over (1 -
\tilde q^n)^{24}} \sim {1\over \tilde q}.
\end{equation}
This allows us to write
\begin{equation}\label{omegaasy}
    d (N) \sim {1\over 2\pi i} \int \left({\beta \over
2\pi}\right)^{12} e^{\beta N + {4\pi^2 \over \beta}} d\beta.
\end{equation}
This integral can be evaluated easily using saddle point
approximation. The  function in the exponent is $f(\beta) \equiv
\beta N + {4\pi^2 \over \beta}$ which has a maximum at
\begin{equation}
f^{\prime} (\beta) = 0 \quad  \textrm{or} \quad  N - {4\pi^2 \over
\beta_c} = 0 \quad \textrm{or} \quad  \beta_c = {2\pi \over
\sqrt{N}}.
\end{equation}
The value of the integrand at the saddle point gives us the leading
asymptotic expression for the number of states
\begin{equation}\label{finalomega}
    d (N) \sim \exp{(4\pi \sqrt{N})}.
\end{equation}
This implies that the ensemble of such BPS  states of a given charge vector $Q$
has  nonzero statistical  entropy that goes to leading order as
\begin{equation}\label{microentropy}
    S_{stat}(Q) := \log (d(Q))= 4\pi \sqrt{Q^{2}/2}.
\end{equation}
We would now like to identify the black hole solution
corresponding to this state  and test if this microscopic entropy
agrees with the macroscopic entropy of the black hole.

The formula that we derived for the degeneracy $d(N)$ is
valid more generally in  any $1+1$ CFT. In a general CFT, the partition
function is a modular form of weight $-k$
\[
Z(\beta) \sim Z \left({4\pi^2 \over \beta}\right) \beta^k.
\]
which allows us to determine high temperature asymptotics from low
temperature asymptotics for $Z (\tilde\beta)$ once again because
\begin{equation}\label{asymp}
     \tilde\beta \equiv {4\pi^2 \over \beta} \to \infty \qquad \textrm{
as} \quad  \beta \to 0.
\end{equation} At low temperature only ground state contributes \beq
Z (\tilde\beta) &=& \Tr \exp({-\tilde\beta (L_0 - c/24)}) \\
&\sim& \exp({-E_0\tilde\beta}) \sim \exp({{\tilde\beta c \over
24}}),\eeq where $c$ is the central charge of the theory. Using
the saddle point evaluation as above we then find.
\begin{equation}\label{cftomega}
     d(N) \sim \exp{({2\pi \sqrt{cN \over 6}})}.
\end{equation}
In our case, because we had $24$ left-moving bosons, $c=24$, and
then (\ref{cftomega}) reduces to (\ref{finalomega}).

\chapter{Spectrum of Quarter-BPS Dyons \label{Quarter}}

In this chapter we  consider the spectrum of quarter-BPS dyons in the simplest string compactification with $\mathcal{N} =4$ in four spacetime dimensions. Surprisingly, the partition function for counting these dyons turns out to involve interesting mathematical objects called Siegel modular forms which are a natural generalizations for  the group  $Sp(2, \mathbb{Z})$   of  usual modular forms of the group  $Sp(1, \mathbb{Z}) \sim SL(2, \mathbb{Z})$.  See {\S\ref{modularbasic}} for  a review of Siegel modular forms and related Jacobi modular forms 

\section{Siegel modular forms and dyons}

Siegel forms occur naturally in the context of counting of
quarter-BPS dyons.  The partition function for these dyons depends on three
(complexified) chemical potentials $(\sigma, \t, z)$, conjugate to the three
T-duality invariant integers 
$$ (P^{2}/2, \, Q^{2}/2, \, P \cdot Q ) :=  (m, n, \ell )$$
respectively and is given by
\begin{equation}\label{igusa}
    \textrm{Z}(\Omega) = \frac{1}{\Phi_{10}(\Omega)} \ .
\end{equation}
Note that this is very analogous to the case of half-BPS states discussed in the tutorials where the partition function was
\begin{equation}\label{igusa}
    \textrm{Z}(\tau) = \frac{1}{\Delta(\tau)} \ .
\end{equation}
was the inverse of a modular form $\Delta(\tau)$ of weight $12$ of the group $Sp(1, \mathbb{Z})$. 

The product representation of the Igusa form is particularly useful for the
physics application because it is closely related to the generating function for
the elliptic genera of symmetric products of $K3$ introduced in the Appendix.  This is a
consequence of the fact that the multiplicative lift of the Igusa form is
obtained starting with the elliptic genus of a single copy $K3$ as the input.
The generating function for the elliptic genera of symmetric
products of K3  is defined by
\be\label{defZ5d}
\widehat{Z}(\s,\t,z) := \sum_{m=-1}^\infty
\chi_{m+1} (\t,z) p^m
\ee
where $\chi_m(\t,z)$ is the elliptic genus of $ \textrm{Sym}^m(K3)$ with $ \chi_{0}(\t, z) \equiv 1$ and $\chi_1(\t,z) \equiv \chi(\t,z)$.
A standard orbifold computation \cite{Dijkgraaf:1996xw} gives
\be\label{ellprzero}
\widehat{Z}(\s,\t,z) = \frac{1}{p}\prod_{s > 0, t \ge 0, r} \frac{1} {(1- p^s q^t y^r)^{C_0(4st- r^{2})}}
\ee
in terms of the Fourier coefficients $C_0$ of the elliptic genus of a single copy of $K3$. As we will explain in the next section, this  partition function captures the degeneracies of bound state of  $m$ D1-branes and  a single D5-brane carrying momentum and spin. 

Comparing the
product representation for the Igusa form (\ref{final2}) with \eqref{ellprzero},
we get the relation:
\be\label{4d5drel}
\textrm{Z}(\Omega) = \frac{1}{\Phi_{10}(\s,\t,z)} = \frac{\widehat{Z} (\s,\t,z)} {\psi(\t,z)} \ .
\ee
This relation of the Igusa form to the elliptic genera of symmetric products of $K3$ and the degeneracies of  D1-D5 bound states has a deeper
physical significance and  allows for a microscopic derivation of the counting formula as we explain below.

The the logic of the derivation is as follows:
\begin{enumerate}
\item We derive the degeneracy for a special charge configuration in one corner of the moduli space.
\item Using constraints from wall-crossing, we extend this answer for the same set of charges to all over the moduli space.
\item Using duality symmetries,  we extend this answer to all possible values of charges. 
\end{enumerate}

With this general strategy in mind, we turn to the derivation of the dyon partition function for a special representative set of charges in a certain weakly coupled region of the moduli space.

\section{A representative  charge configuration}

Consider four-dimensional BPS-states in Type IIB on $K3 \times S^{1} \times \tilde S^{1}$ with the following charge configuration:

\begin{itemize}
\item 1 KK-monopole associated with the circle $\tilde S^{1}$.
\item 1 D5-branes wrapping $K3 \times S^{1}$
\item  $m$ D1-branes wrapping $S^{1}$
\item $n$ units of momentum along the circle $S^{1}$
\item $l$ units of momentum  along the circle $\tilde S^{1}$
\end{itemize}

We would like to compute $d(m, n, l)$ which is the number of quantum states with these quantum numbers counting bosons with +1 and fermions with -1.  Let $F$ be the spacetime fermion number then we could try to compute 
\begin{equation}
 {\textrm{Tr}_{m, n, l}}{\left[ (-1)^{F} \right]} \, .
\end{equation}
However, this vanishes. If a state breaks $2n$ supersymmetries, then it has $2n$ real fermion zero modes which are the Goldstinoes of the broken symmetry.  Quantization of each pair leads to Bose-Fermi degeneracy so the trace above vanishes. This can be remedied by inserting $(2h)^{n}$ where $h$ is the `helicity', that is, the third component of angular momentum in the rest frame. 
For states paired by a complex fermion the effect of this insertion is to `soak up' the fermion zero mode since this mode has spin half. Thus, we compute
\begin{equation}
 d(m, n, l) = \textrm{Tr}_{m, n, l}{\left[ (-1)^{F}(2h)^{6} \right]} \, 
\end{equation}
since for  a quarter-BPS state, out of the $16$ supersymmetries $12$ are broken. In practice, this means we just ignore the $12$ fermionic zero modes from broken supersymmetry and evaluate simply $\textrm{Tr}(-1)^{F}$ over the remaining modes. 
The index thus defined receives contribution only from the BPS states.

It turns out that we can relate these unknown degeneracies  $d(m, n, l)$ of 4d-states  to known degeneracies of the  D1-D5-P configuration in five dimensions which are much easier to compute. This is known as the 4d-5d lift
\cite{Gaiotto:2005xt}. The main idea is to use the fact that the geometry of the
Kaluza-Klein monopole \eqref{tnutgeom} in the charge configuration above asymptotes to $\mathbb{R}^{3} \times \tilde S^{1}$ at asymptotic infinity $r \rightarrow \infty$ but reduces to 
flat Euclidean space $\mathbb{R}^{4}$ near the core of the monopole at $r \rightarrow 0$.  Thus at asymptotic infinity we have a KK-monopole in four-dimensional flat Minkowski spacetime which near the core looks like a  five-dimensional flat Minkowski spacetime.  Our charge configuration  then reduces
essentially to the \textit{five-dimensional} Strominger-Vafa black hole \cite{Strominger:1996sh} with
angular momentum \cite{Breckenridge:1996is} discussed in the previous subsection.

Our strategy will  be to compute the grand canonical partition function introducing chemical potentials $(\s, \t, z)$ conjugate to the charges $(m, n, l)$  and the `fugacities' 
\begin{equation}
p:= e^{2\pi i \s} \, , \quad q:= e^{2\pi i \t} \, , \quad y:= e^{2\pi i z}\, .
\end{equation}
The partition function is then 
\begin{equation}
Z(\s, \t, z) = \sum_{m, n, l} p^{m} q^{n} y^{l } (-1)^{l}\, d(m, n, l) \, .
\end{equation}
The factor of $(-1)^{l}$ is introduced for convenience which can be absorbed by $z \rightarrow z + 1/2$.

Since $d(m, n, l)$ is a topological quantity protected from quantum corrections, 
the dyon partition function it does not depend on the coupling or the moduli such as the radius $\tilde R$. 
We can focus on the region near the core by taking the radius of the
circle $\tilde S^1$ goes to infinity so that in this limit we have a weakly coupled problem. In this limit, the charge $l$ corresponding to
the momentum around this circle gets identified with the angular momentum $l$ in
five dimensions. 
The total partition function at weak coupling at large radius $\tilde R$ is thus a product of three factors
\begin{equation}
Z(\Omega) = Z_{D1}(p, q, y) \, Z_{KK}(q) \, Z_{CM}(q, y) \, .
\end{equation}
The three factors arise as follows.

\begin{myenumerate}
\item The factor $Z_{D1}(\s, \t, z)$ counts the bound states of the D1-brane  bound to a single D5-brane, carrying arbitrary momentum and angular momentum.
\item The factor $Z_{KK}(\tau)$ counts the  bound states of
momentum $n$ with the Kaluza-Klein monopole. The KK-monopole cannot carry any momentum along the $\tilde S^{1}$ directions nor does it carry any D1-brane charge. Hence the partition function depends only $\t$. 
\item  The factor $Z_{CM}(\tau, z)$ counts the bound states of  the center of mass motion of
the Strominger-Vafa black hole in the Kaluza-Klein geometry \cite{Gaiotto:2005hc,
David:2006yn}.  It carries no D1-brane charge and hence depends only $\tau$ and $z$. 
\end{myenumerate}

At weak coupling, these three systems reduce to decoupled bosonic and fermionic oscillators and our computation is reduced to something very similar to perturbative calculation described in the previous chapter. Each oscillator carries certain quantum numbers $(s, t, r)$ which can contribute to the total charge $(m, n, l)$ of our interest. Each bosonic oscillator contributes
\begin{equation}
\sum_{k=0}^{\infty} e^{2\pi i k ( s\s, t \t, r z)} = \left(1 - p^{s}q^{t} y^{r}\right)^{-1} \, .
\end{equation}
Each fermionic oscillator contributes
\begin{equation}
\sum_{k=0}^{1} e^{2\pi i k ( s\s, t \t, r z)} (-1)^{k}= \left(1 - p^{s}q^{t} y^{r}\right) \, 
\end{equation}
where the $(-1)^{k}$ is present because of $(-1)^{F}$. 
The partition function will be thus of the general form
\begin{equation}
Z(\Omega) \sim \prod_{s, t, r} \frac{1} {(1- p^s q^t y^r)^{f(s, t, r)}} \, ,
\end{equation}
where $f(s, t, r)$ is the difference between the number of bosonic oscillators  and the number of fermionic oscillators for given charges $(s, t, r)$ . All physics is now contained in these numbers.
In the remaining subsections we  discuss systematically various contribution to the partition function to determine $f(s, t, r)$ for our system.

\section{Bound states of D1-branes and D5-branes}

As a warm up, let us first consider D1-brane (or fundamental Type-II string) in flat space wrapped around a circle $S^{1}$ or radius $R$ with coordinate $y \sim y + 2\pi R$.
The fluctuations of the D1-brane consists of $8$ transverse bosons
$\phi^{i}(t, y)$ as well as $8$ left-chiral  fermions
$S^{a}(t + y)$ and $8$  right-chiral fermions $\tilde S^{a}(t-y)$ where $t$ is the time coordinate, $i = 1, \ldots, 8$, and $a = 1, \ldots, 8$. These constitute the field content of the 1+1 D CFT living on $S^1$.
The fluctuations are of the form
\begin{equation}
\phi^{i}(t, y) = \phi_{0}^{i} + p_{0}^{i} t + \sum_{n>0}\phi^{i}_{n}e^{-\frac{n}{R}(t -y)} + \sum_{n>0}{\tilde \phi}^{i}_{n}e^{-\frac{n}{R}(t + y)} + c. c.
\end{equation}
For the fermions we have similarly
\begin{eqnarray}
S^{a} (t - y) &=& \sum_{n>0}S^{a}_{n}e^{-\frac{n}{R}(t -y)}  + c. c. \\
{\tilde S}^{a} (t + y) &=& \sum_{n>0}{\tilde S}^{a}_{n}e^{-\frac{n}{R}(t +y)}  + c. c.
\end{eqnarray}
We can quantize this system as usual. Then ${ \phi}^{i}_{n}$
 and ${\tilde \phi}^{i}_{n}$ are bosonic oscillators with frequencies $n/R$ and occupation numbers 
${N}^{i}_{n}$ and ${\tilde N}^{i}_{n}$ respectively. Similarly,  ${ S}^{a}_{n}$ and ${\tilde S}^{a}_{n}$ are fermionic oscillators with frequencies $n/R$ and occupation numbers 
${M}^{i}_{n}$ and ${\tilde M}^{i}_{n}$ respectively. The total left-moving momentum along $S^{1}$ is
\begin{equation}
P = \frac{1}{R}\sum_{i=1}^{8} \sum_{n=1}^{\infty} n (N^{i}_{n} - {\tilde N}^{i}_{n} ) + \frac{1}{R}\sum_{a=1}^{8} \sum_{n=1}^{\infty} n (M^{a}_{n} - {\tilde M}^{a}_{n} ) 
\end{equation}
and the total energy is
\begin{equation}
E =  \frac{1}{R}\sum_{i=1}^{8} \sum_{n=1}^{\infty} n (N^{i}_{n} + {\tilde N}^{i}_{n}) +  \frac{1}{R}\sum_{a=1}^{8} \sum_{n=1}^{\infty} n (M^{a}_{n} + {\tilde M}^{a}_{n} ) 
\end{equation}
To obtain a BPS state we want to minimize the energy given fixed momentum $P$. This implies
\begin{equation}
{\tilde N}^{i}_{n} = 0 \, ,\quad {\tilde M}^{i}_{n} = 0  \quad E =P \, .
\end{equation}
We would like to know how many BPS states there are for a given  charge $P$. This is a combinatorial problem of finding $d(P)$ which is the number of ways to choose a set of integers $\{N^{i}_{n}, M^{a}_{n}\}$ satisfying the constraint
\begin{equation}
 \frac{1}{R}\left( \sum_{n=1}^{\infty} \left( \sum_{i=1}^{8} n N^{i}_{n} +   \sum_{a=1}^{8} \ n (M^{a}_{n} \right) \right) = P \, .
\end{equation}
As usual it is easier to pass to the canonical ensemble.  computing
\begin{equation}
Z(\tau) := \sum_{\{N^{i}_{n}, M^{a}_{n}\}} q^{N} \equiv \sum_{P} d(N) q^{N}\, , \quad q:=e^{2\pi i \t } \, ,
\end{equation}
ignoring the constraint. Here we have use for convenience $N = RP$ which is an integer or  equivalently can absorb $R$ into $\tau$.  One can then obtain $d(N)$ by inverse Laplace transform using\begin{equation}
Z(\tau) := \sum_{P} d(N) q^{N} \, ,\quad d(N) = \int_{0}^{1} e^{-2\pi i N \tau} Z(\tau) d\t \, .
\end{equation}
The partition function is readily evaluated and is given by
\begin{equation}
Z(\tau) = \frac{\prod_{n=1}^{\infty}( 1 + q^{n})^{8}}{\prod_{n=1}^{\infty}( 1 - q^{n})^{8}}
\end{equation}
{From} this one can find that 
\begin{equation}
d(N) \sim e^{2\pi \sqrt{2N}} \, ,
\end{equation}
which follows also from the Cardy formula applied to the worldsheet CFT living on the circle, using the fact that for $8$ free bosons and $8$ free fermions the central charge is $12$.

After this warm-up exercise, let us turn to the problem of motion of $m$ D1-branes bound to a single D5-brane. Now, \textit{a priori} the D1-brane can again oscillate in all $8$ transverse directions. However, if we switch on a 2-form field along 2-cycles of $K3$, then open strings connecting D1-branes and D5-branes become tachyonic. Condensation into ground state binds the D1-branes to the D5-branes and as a result they can oscillate only along the directions along the $K3$. 

We are interested in a configuration with $m$ units of D1-brane charge $n$ units of momentum, and $l$ units of angular momentum. If $m$ is divisible by $s$ then we have to consider both the configuration with $m$  D1-branes winding  number $1$ as well as the configuration with $m/s$ D1-branes with winding number $s$. Similarly, the momentum and angular momentum can be shared among these $m$ or $m/s$ D1-branes. As usual, it is more convenient to relax all constraints on the charges and compute instead the (grand) canonical partition function. So,  we introduce chemical (complexified) chemical potentials $\sigma, \tau, z$ conjugate to the integers $m, n, l$ and compute the unrestricted sum   by summing over all possible charges $(r, s, t)$. The degeneracies $d_{D1}(m, n, l)$ can then be extracted by an inverse Fourier transform.

Consider a D1-brane wound $r$ times along the $S^{1}$, carrying momentum $s$ along the $S^{1}$ with angular momentum $J_{L} = t/2$.  Let 
\begin{equation}
Z_{D1} = \frac{1}{p}\prod_{s > 0, t \ge 0, r} \frac{1} {(1- p^s q^t y^r)^{c(s, t, r)}} \, .
\end{equation}
Now, a D1-brane wrapping $s$ times around a circle $R$ is  like a D1-brane wrapping once   on a circle of  effective radius  $R_{e}= 2\pi R s$. If we want it to carry physical momentum $t$, then since
\begin{equation}
\frac{t}{R} = \frac{t s}{nR} = \frac{t s}{R_{e}}
\end{equation}
Because of conformal invariance, the partition function does not depend on the overall scale $R$. We thus conclude that the partition function for winding $s$ and physical momentum $t$ is the same as the partition function for winding $1$ and physical momentum $st$. In other words,
\begin{equation}
c(s, t, r) = c_{0}(st, r) \, .
\end{equation}
These coefficients are  nothing but the $c_{0}(n, l)$ defined in \eqref{phizeroFE} of the elliptic genus $\chi(\t, z)$ of a single copy of $K_{3}$. Hence $c(s, t, r) = c_{0}(st, r)  = C_{0}(4st-r^{2})$  from  \eqref{defC}. Indeed, our computation of $Z_{D1}$ is one way to derive the generating function $\hat Z$ for the elliptic genera of symmetric products of $K3$.
In summary,
\begin{equation}
Z_{D1}(\sigma, \tau, z) = \hat Z(\sigma, \tau, z) \, .
\end{equation}

\textbf{\it Comment:}
The problem of counting microstates of $m$ D1-branes bound to a D5-brane is  the counting problem that arises in computing the microstates of the 
well-known  Strominger-Vafa black hole in five dimensions \cite{Strominger:1996sh}. The 
microscopic configuration there consists of $Q_5$ D5-branes wrapping
$K3\times S^1$, $Q_1$ D1-branes wrapping the $S^1$, with total
momentum $n$ along the circle. We have chosen $Q_{5}=1$ and $Q_{1}=m$ but more generally, we can simply replace $m$ by $Q_{1}Q_{5}$. The bound states are described by
an effective string wrapping the circle carrying left-moving
momentum $n$. The central charge of the system can be computed at
weak coupling and is given by $6m$. In this system, the leading order entropy at large charge can be computed by applying the Cardy formula provided we operate in a certain regime in moduli and charge space.
 We work in a region of moduli space where the K3 is small compared to the S1. In such a situation, the dynamics of the D1-D5 system are encapsulated in a 1+1 D CFT living on $S^1$. The D1-D5-P configuration can then be regarded as a state in this CFT with the right moving oscillators fixed to their ground state and the left moving excitation number or CFT temperature  proportional to  $n$. Then in the limit of $n \gg Q1\,Q5$, the Cardy formula for the high temperature expansion of the CFT can be used to compute the leading order degeneracy of the state. 
Applying Cardy's
formula therefore, gives,
\begin{equation}\label{omegatwo}
    d_{m} (n) = \exp ({2\pi \sqrt{{mn}}}) .
\end{equation}
 This  implies a microscopic entropy $S = \log{d} =  2\pi
\sqrt{Q_1Q_5 n}$. The corresponding BPS black hole solutions with
three charges in five dimensions can be found in supergravity and
the resulting entropy matches precisely with the macroscopic
entropy \cite{Strominger:1996sh}.

\section{Dynamics of the KK-monopole}

In the previous subsection we have worked out the low-energy massless fluctuations of the KK-monopole. If we excite only the left-movers then we have 
$24$ bosons carrying momentum $t$. The KK-monopole cannot support any momentum along $the S^{1}$ circle. Summing over all momenta gives rise to the partition function
\begin{equation}
Z_{KK}(\tau) = {1 \over q} \prod_{t=1}^\infty {1\over(1 -
  q^t)^{24}} = \frac{1}{\eta^{24}(\t)}\quad 
\end{equation}
The factor of $1/q$ comes because the ground state carries some `zero point' momentum $-1$. Altogether, we recognize this as precisely the partition function of the left-moving BPS oscillations of the heterotic string as expected from duality.

\section{D1-D5 center-of-mass oscillations}

Now it remains for us to find the contribution to the partition function from the oscillations of the center of mass of the D1-D5 system moving in the background the KK-monopole. This is easy to evaluate using the fact that for large radius near the center of the KK-monopole, the Taub-NUT space is essentially flat Euclidean space $\mathcal{R}^{4}$.
The partition function of four bosons and four fermions is simply
\begin{equation}
Z_{CM}(\tau, z) = \frac{\eta^{6}(\tau)}{\theta_{1}^{2}(\tau, z) } \, .
\end{equation}

Putting this all together we find the desired result
\begin{equation}
Z(\Omega) = \frac{\hat Z(\s, \t, z)}{\psi(\tau, z)} = \frac{1}{\Phi_{10}(\Omega)} \, .
\end{equation}

\section{Wall-crossing and contour prescription}

Given the partition function \eqref{igusa}, one can extract the black hole
degeneracies from the Fourier coefficients. However, there is one complication that also turns out to have interesting physical implications. The Igusa cusp form has double zeros at $z=0$ and its images. The partition
function is therefore a \textit{meromorphic} Siegel form (\ref{trans2}) of weight
$-10$ with double poles at these divisors. As a result, different Fourier contours would give different answers for the degeneracies and there appears to be an ambiguity in the choice of the Fourier contour. 

This ambiguity turns out  to have  a very nice physical interpretation. The spectrum of quarter-BPS dyons actually has a moduli dependence. For a given charge vector $\Gamma$, there are single-centered black hole solutions that exist everywhere in the moduli space. However, in addition, there can be two-centered solutions such that  one center carries charge $\Gamma_{1}$ and the other $\Gamma_{2}$ with $\Gamma = \Gamma_{1}+ \Gamma_{2}$. 
A simple example is when one charge center has charge $(Q, 0)$ and the other has charge $(0, P)$. The distance between these two centers is fixed in terms of the charges and the moduli fields. 

As one changes the moduli, the distance between the two centers can go to infinity and the two-centered solution can decay at certain walls \textit{i.}\textit{e.} surfaces of co-dimension one. Thus, on one side of the wall, we have only a single-centered black hole whereas on the other side we have the single-centered black hole as well as the two-centered black hole. Hence the degeneracy on one side of the wall is different from the degeneracy on the other side of the all. Upon crossing the wall, the degeneracy jumps. This phenomenon is known as the `wall- crossing phenomenon'. 
The moduli space is thus divided up into chambers separated by walls. The degeneracy is different from chamber to chamber. 

This dependence of the degeneracy on the chamber in the moduli space is nicely captured by the dependence of the Fourier coefficients on the choice of the contour. 
 As we will explain below, the choice of the contour depends on the moduli in a precise way. As the moduli are varied, the contour is deformed. The dependence of the contour on the moduli is such that as the moduli hit a wall in the moduli space, the contour hits a pole of the partition function. 
The poles are thus nicely correlated with the walls. Crossing the wall in the moduli space corresponds to crossing a pole in the contour space. The jump in the degeneracy upon crossing the wall is given by the residue at the pole that is crossed by the contour. 

To see this more precisely, note that the three quadratic T- duality
invariants of a given dyonic state can be organized as a $2 \times 2$ symmetric
matrix
\be\label{matrix_charge_vector}
\L  =
\left(
\begin{array}{cc}
 Q\cdot Q & Q\cdot P\\
    Q\cdot P & P\cdot P
    \end{array}
\right) \, =
\left(
\begin{array}{cc}
 2n& \ell \\ 
 \ell & 2 m 
\end{array}
\right)
\; ,
\ee
where the dot products are defined using the $O(22, 6; \mathbb{Z})$ invariant
metric L. The matrix $ \Omega$ in (\ref{igusa}) and (\ref{period}) can be viewed
as the matrix of complex chemical potentials conjugate to the charge matrix $\L$.
The charge matrix $\L$ is manifestly T-duality invariant. Under an S-duality transformation (\ref{Sgroup}),  it transforms as
\begin{equation}\label{lambdatransform}
    \L \to \gamma \L \gamma^t
\end{equation}
There is a natural embedding of this physical S-duality group $SL(2, \mathbb{Z})$
into $Sp(2, \mathbb{Z})$:
\begin{equation}\label{sembed}
\left(
  \begin{array}{cc}
    A & B \\
    C & D \\
  \end{array}
\right)
   = \left(
      \begin{array}{cc}
        (\g^t)^{-1} & \textbf{0} \\
        \textbf{0} & \g   \\
      \end{array}
    \right)    =
    \left(
      \begin{array}{cccc}
        d & -c & 0 & 0 \\
        -b & a & 0 & 0 \\
        0 & 0 & a & b \\
        0 & 0 & c & d \\
      \end{array}
    \right)
   \, \in  Sp(2, \mathbb{Z}).
\end{equation}
The embedding is chosen so that $\Omega \to (\g^T)^{-1} \Omega \g^{-1}$ and $\Tr
(\Omega \cdot \Lambda)$ in the Fourier integral is invariant. This choice of the
embedding ensures that the physical degeneracies extracted from the Fourier
integral are S-duality invariant if we appropriately transform the moduli at the
same time as we explain below.

To specify the contours, it is useful to define the
following moduli-dependent quantities. One can define the matrix of right-moving
T-duality invariants
\be\label{rightmatrix_charge_vector}
\L_{R}  =\bem Q_R\cdot Q_R & Q_R\cdot P_R \\
    Q_R\cdot P_R & P_R\cdot P_R \eem
\;.
\ee
which depends both on the integral charge vectors $N, M$ as well as the T-moduli $\mu$
One can then define two matrices naturally associated to the S-moduli $\l = \l_{1 }+ i\l_{2}$
and the T- moduli $ \mu$ respectively by
\be\label{ST_moduli}
\mathcal{S} = \frac{1}{\l_{2}} \bem |\l|^2 & \l_{1}\\\ \l_{1}& 1 \eem\quad,
\quad \mathcal{T} =
\frac{ \L_{R}}{{|\det
(\L_{R})|^\frac{1}{2}}}\;.
\ee
Both matrices are normalized to have unit determinant. In terms of them, we
can construct the moduli-dependent `central charge matrix'
\be\label{def_Z_vec}
{\mathcal Z} = {{|\det(\L_{R})|^\frac{1}{4}}}\, \big(\mathcal{ S}+\mathcal{ T}
\big) ,
\ee
whose determinant equals the BPS mass
\be
M_{Q,P} = |\det{\mathcal Z}|.
\ee
We define 
\begin{equation}
\tilde \Omega \equiv
     \left( \begin{array}{cc} \sigma &  -z \\  -z & \tau \\ \end{array} \right)
\end{equation}
related to $\Omega$ by an $SL(2, \mathbb{Z})$ transformation
\begin{equation}
\tilde \Omega = \hat S \Omega \hat S^{-1}  \quad where \quad \hat S=    \left( \begin{array}{cc} 0 &  1 \\  -1 & 0 \\ \end{array} \right)
\end{equation}
so that, under a general S-duality transformation $\gamma$, we have the transformation $\tilde \Omega \rightarrow \gamma \tilde \Omega \gamma^{T}$ as  $\Omega \to (\g^T)^{-1} \Omega \g^{-1}$.

With these definitions,  $\Lambda, \Lambda_{R}, \mathcal{Z}$ and $\tilde \Omega$ all transform as $X \to \g
X \gamma^T$ under an S-duality transformation (\ref{Sgroup}) and are invariant under T-duality transformations. 
The moduli-dependent Fourier contour can then be  specified in a duality-invariant fashion by
\cite{Cheng:2007ch}
\be\label{contour}
\mathcal{C} = \{\im \tilde\O = \varepsilon^{-1} \mathcal{Z};\quad 0 \leq \Re(\t),
\Re(\s), \Re(z) < 1 \},
\ee
where $\varepsilon \rightarrow 0^+$. For a given set of charges, the contour
depends on the moduli $\l, \mu$ through the definition of the central
charge vector (\ref{def_Z_vec}). The degeneracies $d(m, n, l)\lvert_{\l, \m}$ of
states with the T-duality invariants $ (m, n, l)$, at a given point $(\l,
\mu)$ in the moduli space are then given by\footnote{The physical degeneracies have an additional multiplicative factor of  $(-1)^{\ell +1}$ which we omit here for simplicity of notation in later chapaters.} 
\begin{equation}\label{inverse}
   d(m, n, l)\lvert_{\l, \m} = \int_{\mathcal{C}}
   \, e^{-i\pi \Tr (\Omega \cdot \Lambda)}\, \,
   {\textrm{Z}(\Omega)} \, d^3 \Omega \, .
\end{equation}

This contour prescription thus specifies how to extract the degeneracies from the partition function for a given set of charges and in any given region of the moduli space. In particular, it also completely summarizes all wall-crossings as one moves around in the moduli space for a fixed set of charges. 
Even though the indexed partition function has the same functional form
throughout the moduli space, the spectrum is moduli dependent because of the
moduli dependence of the contours of Fourier integration and the pole structure
of the partition function. 
Since the degeneracies depend on the moduli \textit{only} through the dependence
of the contour $\mathcal{ C}$, moving around in the moduli space corresponds to
deforming the Fourier contour. 

With this understanding of the wall crossing and the contour prescription, we have completely specified how to extract dyon degeneracies from the Fourier coefficients of the partition function. The partition function in turn is constructed explicitly in terms of Fourier coefficients of known objects such as $\psi$ or $\chi$. We will not here analyze wall-crossing in any further detail which can be found in \cite{Dabholkar:2007vk,Sen:2007vb, Cheng:2007ch}.

\section{Asymptotic expansion \label{Asym}}

Given the exact formula for the degeneracies, one can try to extract the asymptotic degeneracies in the limit where $m, n$ are both large and positive. Since the Fourier integral now involves three variables, the  calculation is more involved than the Cardy formula that we encountered for modular forms of  single variable. The answer however is simple. The statistical entropy $\log (d )$ is obtained by minimizing the following function with respect to $\lambda$
\be\label{einfo4}
\mathcal{E}_B(\l) = {\pi\over 2 \l_{2}} \, |Q +\lambda P|^2
-64\pi^{2}\phi(\lambda, \bar \lambda) + O(Q^{-2})\, ,
\ee
where $\phi(\l, \bar \l)$ :
\begin{equation}\label{correction-function}
\phi(\l, \bar \l) = -\frac{1}{64\pi^{2}}\left\{ 12 \log \left[-2i (\l -\bar \l)\right]+ 24 \log\left[ \eta(\l)\right] + 24 \log\left[\eta(\bar \l)\right] \right\} \, .
\end{equation}
For a detailed description of the expansion, see \cite{LopesCardoso:2004xf, David:2006yn}.

\chapter{Quantum Black Holes \label{Quantum}}

Now we turn to the black holes in string theory that corresponds to the ensembles of the BPS quantum microstates.  Such dyonic BPS black holes are essentially generalizations of the Reissner-Nordstr\"om black hole  but now with both electric and magnetic charges under several different $U(1)$ gauge fields.  They are solutions of the effective action of string theory which contains many more terms compared to the Einstein-Maxwell action \eqref{action}.

To view a  black hole as an ensemble of states, it is important to find  the black hole solution of the full effective action that connects the near horizon region that we analyze below  to  an asymptotically flat spacetime. For the leading two-derivative effective action of toroidally compactified heterotic string theory, such exact interpolating solutions  for dyonic BPS black holes  are known  \cite{Sen:1994eb, Cvetic:1995uj}. The black hole geometry exhibits the attractor mechanism: the values of scalar fields get `attracted' to their atttactor values at the horizon that are determined entirely by the charges of the black hole and independent of their values at  asymptotic  infinity \cite{Ferrara:1995ih, Ferrara:1996dd, Strominger:1996kf}.  
Incorporating the effect of higher-derivative terms in the effective action for the interpolating solutions is in general much more complicated  and can be found in \cite{LopesCardoso:1998wt,LopesCardoso:1999cv,LopesCardoso:1999xn,LopesCardoso:2000qm}. 

For our purposes, we are only interested in the near-horizon properties of the black hole such as its entropy and the attractor values of various scalar fields at the horizon. 
This can be analyzed much more simply using the entropy function formalism developed in \S\ref{EntropyFunction}.

 In section \S\ref{Leading} we discuss the near horizon solution and the entropy for the leading two-derivative effective action and consider the correction to the Wald entropy to the next subleading order in \S\ref{Subleading}.  They compare beautifully with statistical entropy given by the logarithm of  the microscopic degeneracies computed in the \S\ref{Quarter}.

The case of black holes corresponding to the half-BPS states is in some ways more interesting which we discuss in section \S\ref{Small}. In this case, the entropy is actually zero to leading order because the geometry has a null singularity instead of a smooth horizon. The area of the event horizon is thus zero to leading order.  Subleading quantum corrections modify the geometry so that the corrected geometry has a string scale horizon. The Wald entropy associated with this horizon precisely matches with the statistical entropy computed in \S\ref{Half}.

\section{Wald entropy to leading order\label{Leading}}

For a state with electric charge vector $q$ and magnetic charge vector $p$, the fields near the horizon take the form\footnote{For an extensive description of this computation see \cite{Sen:2005iz}.}
\begin{eqnarray}
 ds^2&=&\frac{v_{1}}{16}\left(-(\sigma^2-1)d\t^2+\frac{d\sigma^2}{\sigma^2-1}\right)\,\, + \frac{v_{2}}{16}\left(d\theta^{2} + sin^{2}\theta d\phi^{2}\right)   \\
  F^{(i)}_{\s\t}&=& \frac{1}{4}e_i \, , \quad F^{(i)}_{\theta\phi}= \frac{1}{16\pi}p_{i} \, ,\quad 
 M_{ij} = u_{ij}, \,\, \quad  S=u_s, \,\, \quad a=u_a \, . \nonumber
\end{eqnarray}
Substituting into the action \eqref{hetaction} we get
\begin{eqnarray} \label{eag5}
&& \qquad  f(u_S, u_a, u_M, \vec v, \vec e, \vec p)
\equiv \int d\theta d\phi \, \sqrt{-\det G} \, \mathcal{L}
 \\ 
&=& 
{1\over 8} \, v_1 \, v_2
\, u_S  \left[ -{2\over v_1} +{2\over v_2} + 
 {2\over v_1^2} e_i (Lu_M L)_{ij} e_j - {1\over 8\pi^2 v_2^2}  p_i 
 (Lu_ML)_{ij} p_j + { u_a\over \pi u_S v_1 v_2}  e_i L_{ij} p_j
 \right] \, . \nonumber 
\end{eqnarray}
Hence the entropy function becomes
\begin{eqnarray}\label{eag7pre2}
\mathcal{E}(q, u_S, u_a, u_M, v,  e,
 p) &:=& 2\pi \left( e_i q_i 
- f(u_S, u_a, u_M, v, e, p) \right) \nonumber \\
&=& 2\pi \Bigg[ e_i q_i - {1\over 8} \, v_1 \, v_2
\, u_S  \bigg\{ -{2\over v_1} +{2\over v_2} + 
 {2\over v_1^2} e_i (Lu_M L)_{ij} e_j \nonumber \\
 && \qquad - {1\over 8\pi^2 v_2^2}  p_i 
 (Lu_ML)_{ij} p_j + { u_a\over \pi u_S v_1 v_2}  e_i L_{ij} p_j
 \bigg\} \Bigg]\, .
\end{eqnarray}
Eliminating $e_i$ from 
\eqref{eag7pre} using the equation $\partial \mathcal{E}/\partial e_i=0$ we get:
\begin{eqnarray} \label{eag7}
&&\qquad\qquad \mathcal {E}(q, u_S, u_a, u_M, v, e(u, v, q, p),  p)  = \nonumber
\\  &&2\pi \bigg[ {u_S\over 4} (v_2 - v_1)
+{v_1\over v_2 u_S} \, q^T u_M q 
+{v_1\over 64\pi^2 v_2 u_S} (u_S^2 + u_a^2)
p^T L u_M L p \nonumber  -{v_1\over 4 \pi v_2 u_S}  \, u_a\, q^T u_M L p\bigg]\,.
\end{eqnarray}
We can simplify the formul\ae\ 
by defining new charge vectors:
\be \label{eag8a}
Q_i = 2 q_i, \qquad P_i = {1\over 4\pi}\, L_{ij} p_j\, ,
\ee
which are normalized so that they are integral and satisfy the Dirac quantization condition. 
In terms of $\vec Q$ and $\vec P$
the entropy function $\mathcal{E}$ is given by:
\be \label{eag8b}
\mathcal{E} = {\pi\over 2} \bigg[ u_S(v_2 - v_1) +{v_1\over v_2 u_S} 
\left( Q^T u_M 
Q +
 (u_S^2 + u_a^2) \, P^T u_M P - {2 }
\, u_a \, Q^T u_M P \right)\bigg]\, .
\ee
Substituting \eqref{eag12} into \eqref{eag8b} and using \eqref{eag10},
\ref{eag11}, we get:
\be \label{eag13}
\mathcal{E}= {\pi\over 2} \left[ u_S(v_2-v_1) 
+{v_1\over v_2} \left\{{Q^2\over u_S} + 
{P^2\over u_S} (u_S^2 + u_a^2)
- 2\, {u_a\over u_S}\, Q\cdot P \right\} \right]\, .
\ee
Note that we have expressed the right hand side of this 
equation in an T-duality invariant form.
Written in this manner, eq.\ref{eag13} is valid
for general $\vec P$, $\vec Q$ satisfying 
\be \label{erg1}
P^2>0, \quad Q^2>0, \quad (Q\cdot P)^2<Q^2P^2\, .
\ee
 We now need to find the extremum of $ \mathcal{E}$ with respect to $u_S$, $u_a$,
$u_{Mij}$, $v_1$ and $v_2$. In general this leads to a complicated set
of equations. We can simplify the analysis by using the $O(22, 6; \mathbb{R})$ symmetries  \eqref{ttransform} of the two-derivative action \eqref{hetaction} which induces the following transformations 
on the various parameters:
 \begin{eqnarray} \label{eag9a}
e_i\to \Omega_{ij} e_j, \qquad  p_i\to \Omega_{ij} p_j, \qquad u_M\to
\Omega u_M \Omega^T\, , \nonumber \\
q_i\to (\Omega^T)^{-1}_{ij} q_j \, , \qquad
Q_i\to (\Omega^T)^{-1}_{ij} Q_j, \qquad P_i \to 
(\Omega^T)^{-1}_{ij} P_j\, .
 \end{eqnarray}
The entropy function  \eqref{eag8b} is invariant under these 
transformations.
Since at its extremum with respect to $u_{Mij}$
the entropy function depends only on $\vec P$, $\vec Q$, $v_1$, $v_2$, 
$u_S$ and $u_a$ it must be a function
of  the $O(22, 6)$ invariant combinations:
\be \label{eag10}
Q^2 = Q_i L_{ij} Q_j, \quad P^2 = P_i L_{ij} P_j, \quad 
Q\cdot P = Q_i L_{ij} P_j\, ,
\ee
besides $v_1$, $v_2$,
$u_S$ and $u_a$.
Let us for definiteness take $Q^2>0$, $P^2>0$, 
and $(Q\cdot P)^2<Q^2 P^2$. In that case with the help
of an $SO(22, 6)$ transformation we can make 
\be \label{eag11}
(I_r-L)_{ij}Q_j=0, \quad (I_r-L)_{ij}P_j=0\, ,
\ee
where $I_r$ denotes the $r\times r$ identity matrix.
This
is most easily seen by diagonalizing 
$L$ to the form 
\begin{equation}
\left(
\begin{array}{cc}
 -I_{22} & 0    \\
  0 &  I_{6}    
\end{array}
\right) \, .
\end{equation}
In this case $Q$ and $P$ satisfying  \eqref{eag11}
will have 
\be\label{eqipi}
Q_i=0, \quad P_i = 0, \quad \hbox{for $1\le i\le 22$}\, .
\ee
Let us now see that for $P$ and 
$Q$ satisfying this condition, every
term in  \eqref{eag8b} is extremized 
with respect to $u_{M}$ for
\be \label{eag12}
u_M  = I_r\, .
\ee
Clearly a variation $\delta u_{Mij}$
with either $i$ or $j$ in the range $[7,r]$ 
will give vanishing contribution to each term in
$\delta  \mathcal{E}$ computed from  \eqref{eag8b}. 
On the other hand due to the constraint  \eqref{mconstraints} on $M$, any 
variation $\delta M_{ij}$ (and hence 
$\delta u_{Mij}$) with $1\le i,j\le 6$ must vanish, since in this
subspace  satisfying \eqref{mconstraints} requires $M$ to be both symmetric and 
orthogonal.
Thus each term in $\delta  \mathcal{E}$ vanishes under all
allowed variations of $u_M$.

We should emphasize that \eqref{eag12} is not the only possible value of $u_M$ that
extremizes $ \mathcal{E}$. 
Any $u_M$ related to  \eqref{eag12} by an $O(22, 6)$ transformation
that preserves the vectors $\vec Q$ and $\vec P$ will extremize
$ \mathcal{E}$.
Thus there is  a family of  extrema
representing flat directions of $ \mathcal{E}$. However
as we have argued in \S\ref{Wald}, the value of the entropy is
independent of the choice of $u_M$.

It remains to extremize $\mathcal{E}$ 
with respect to $v_1$, $v_2$, $u_S$ and $u_a$.
Extremization with respect to $v_1$ and $v_2$ give:
\be \label{es10}
v_1=v_2 = u_S^{-2} \, \left({Q^2} +
{P^2} (u_S^2 + u_a^2) - 2 u_a\, Q\cdot P\right)\, .
\ee
Substituting this into \eqref{eag13} gives:
\be \label{es11copy}
\mathcal{E}= {\pi\over 2}  
{1\over u_S}\, \left\{{Q^2} - 2\, {u_a}\, Q\cdot P
+
{P^2} (u_S^2 + u_a^2)\right\} 
\, .
\ee
It is convenient to write it in a manifestly $SL(, \mathbb{Z})$ invariant way as
\be \label{entropyfinal}
\mathcal{E}= {\pi\over 2}  
{1\over \l_{2}}\, |Q + \l P|^{2} 
\, .
\ee
if we write $\l = u_{a} + i u_{S}:= \l_{1} + i \l_{2}$.

Finally, extremizing with respect to $u_a$, $u_S$ we get
\be \label{eag14}
u_S= {\sqrt{Q^2 P^2 - (Q\cdot P)^2}\over P^2} \, ,
\qquad u_a = 
{Q\cdot P\over P^2}, \qquad
v_1=v_2= 2 P^2  \, .
\ee
The black hole entropy, given by the value of $\mathcal{E}$ for this configuration,
is 
\be \label{eag15}
S_{BH} = \pi \, \sqrt{Q^2 P^2 - (Q\cdot P)^2}\, .
\ee

To get an idea about orders of magnitude let us take $Q \cdot P =0$ for simplicity. Then from \eqref{eag15}   the radius $r_{H}$ of the horizon of the black hole scales as 
\be
r_{H}^{2 } \sim \sqrt{Q^{2} P^{2}} \, \ell_{4}^{2}
\ee  
where $\ell_{4}$ four-dimensional planck length. The four dimensional string coupling $g_{4}^{2}$ at the horizon can be read off from the attractor value of the dilaton in \eqref{eag14}:
\be
g_{4}^{2} = \frac{1}{u_{S}} =  \sqrt{\frac{P^{2}}{Q^{2}}} \, .
\ee
We see that string loop corrections are small if $P^{2} \ll Q^{2}$. 
The string length $\ell_{s}$ is related the Planck length by
\be
\ell_{4} = g_{4} \, \ell_{s} \, .
\ee
Hence the $\apm$ corrections are small if the radius curvature is large in string units, that is, if
\be
r_{H}^{2}/\ell_{s}^{2} \sim  P^{2} \gg 1 \, .
\ee
Hence if we take $Q^{2} \gg P^{2} \gg 1$, we  can compute the Wald entropy in a systematic expansion in $1/Q^{2}$ keeping both the $\apm$ and string loop corrections small.

\section{Subleading corrections to the Wald entropy \label{Subleading}}

The asymptotic expansion in \S\ref{Asym} is obtained in the regime when all charges scale the same way and are much larger than one. In other words, 
\be
Q^{2} \sim P^{2} \gg 1 \, .
\ee
We have already computed the leading order entropy for in section \eqref{Leading}. We would now like to see how to take the effects of higher order corrections. Let us suppose the Lagrangian is of the form
\begin{equation}
\mathcal{L} = \mathcal{L}_{0} + \epsilon \mathcal{L}_{1} \, ,
\end{equation}
where the term of order $\e$ is a small correction from higher-derivative terms. The entropy function defined using this Lagrangian will also be of the form
\begin{equation}
\mathcal{E} = \mathcal{E}_{0} + \epsilon \mathcal{E}_{1}  \, .
\end{equation}
The solutions of the extremization equations will also have an expansion
\begin{eqnarray}
e^{*}(q, p) &=& e^{*} _{(0)} + \e e^{*} _{(1)} + \ldots \, ; \nonumber \\
 u^{*}(q, p) &=& u^{*} _{(0)} + \e u^{*} _{(1)} + \ldots \, ;\quad v^{*}(q, p) = v^{*} _{(0)} + \e v^{*} _{(1)} + \ldots \quad . 
\end{eqnarray}
To compute the entropy we have to compute the value of the entropy function $\mathcal{E}^{*}$ at the extermum
\begin{equation}
\mathcal{E}^{*}(q, p) = \mathcal{E}_{0}(q,  u^{*}, v^{*}, e^{*}, p)+ \epsilon \mathcal{E}_{1}(q,  u^{*}, v^{*}, e^{*}, p)   \, .
\end{equation}
If we are interested in the first subleading correction to order $\e$ we simply expand these functions to obtain
\begin{equation}
\mathcal{E}^{*}(q, p) = \mathcal{E}_{0}(q,  u^{*}_{0}, v^{*}_{0}, e^{*}_{0}, p)+ \epsilon \mathcal{E}_{1}(q,  u^{*}_{0}, v^{*}_{0}, e^{*}_{0}, p)   + O(\e^{2}) \, .
\end{equation}
The important point is that to $O(\e)$ one could have had terms like
\begin{equation}\label{extremum0}
{\partial \mathcal{E}_{0}\over \partial e}\, , \quad 
{\partial \mathcal{E}_{0}\over \partial v}\, ,  \quad  
{\partial  \mathcal{E}_{0}\over \partial  u}\, ,
\end{equation}
evaluated at the leading order extremum values $u^{*}_{0}, v^{*}_{0}, e^{*}_{0}$. However, these all vanish because to the leading order, the extremum values of near horizon fields are found precisely by setting all terms in \eqref{extremum0}  to zero. Hence, to find the first subleading correction, it is not necessary to solve the extermization equations all over again. It suffices to evaluate the correction to the entropy $\mathcal{E}_{1}$ at the extremum values found using the zeroth order entropy function $\mathcal{E}_{0}$. This greatly simplify practical computations. 

To illustrate these ideas, we apply them  to the heterotic action for the dyonic black holes of our interest. The heterotic supergravity action \eqref{hetaction} is only the leading 2-derivative supergravity approximation to the full string effective action. The theory has  a 4-derivative correction to the effective action given by the lagrangian
\begin{equation}\label{corr}
\Delta\mathcal{L} = \phi(\l, \bar \l)\left(R_{\mu\nu\a\b}R^{\mu\nu\a\b} - 4 R_{\mu\nu} R^{\mu\nu}\right) \, ,
\end{equation}
where $\phi(\l, \bar \l)$ is a nontrivial function of axion-dilaton $\l := a +i S$:
\begin{equation}\label{correction-function2}
\phi(\l, \bar \l) = -\frac{1}{64\pi^{2}}\left[ 12 \log(S) + 24 \log\left( \eta(a -i S)\right) + 24 \log\left( \eta(a +i S)\right) \right] \, .
\end{equation}
Note that this is exactly the same function $\phi(\l, \bar\l)$ introduced in \eqref{correction-function}. 
It is easy to check that addition of this term  induces a correction to the entropy function of the form
\begin{equation}
\mathcal{E}_{1} = 64 \pi^{2}\phi(\l, \bar \l) \, .
\end{equation}
Consequently, the  Wald entropy corrected to this order is then given by
\begin{equation}\label{finalwald}
S_{wald} = \pi\sqrt{Q^{2} P^{2} -(Q\cdot P)^{2}} + 64 \pi^{2}\phi\left(a=\frac{Q\cdot P}{P^{2}} \, , \,S = \frac{\sqrt{Q^{2} P^{2} -(Q\cdot P)^{2}}}{P^{2}}\right)  + \ldots
\end{equation}

As a result,  the thermodynamic Wald entropy given by \eqref{finalwald} matches beautifully with the statistical entropy given by \eqref{einfo4} not only to the leading order but also the next subleading order. As mentioned in the preface, the subleading finite size corrections have much more structure than the leading Bekenstein-Hawking entropy and involve a rather nontrivial modular function $\phi$. 

We should emphasize that the origin of this function  in the two computations is of totally different. In the computation of the Wald entropy $S_{wald}(Q, P)$, it arises from specific terms in the effective action of {massless} fields in string theory. In the computation of the statistical entropy $\log( d(Q, P))$, on the other hand, it arises  from the asymptotic expansion of the Fourier coefficients of the partition function for quarter-BPS dyons which for some reason is related the Igusa cusp form. This thus points to a highly nontrivial internal consistency in the structure of string theory and gives us some confidence that we may be on the right track in the search for a quantum theory of gravity.

\section{Wald Entropy of small black holes \label{Small}}

For half-BPS black holes, we can choose a duality frame in which they are purely perturbative with electric charge vector $Q$ and no magnetic charge, or $P=0$.  
In this case, it follows from \eqref{eag14} and \eqref{eag15} that the near horizon solution of the  leading order two derivative action is singular.  In particular, the area of the horizon goes to zero and the attractor value of the string coupling constant goes to zero. 
Thus, in this case it is not sensible to study the effects of higher derivative terms as small corrections to the leading order solution. Rather,  one must consider the full entropy function and find the near horizon geometry by extremizing it.  It turns out that  upon the inclusion of  $\apm$ corrections, the near horizon geometry is no longer singular but has   a horizon with area of order one in  string units. Such black holes with a small string scale horizon have been termed `small' black holes \cite{Dabholkar:2005by, Dabholkar:2005dt}.  Moreover, the Wald entropy of this horizon precisely agrees with the statistical entropy \cite{Dabholkar:2004yr, Dabholkar:2004dq}. This is an interesting phenomenon which illustrates that quantum corrections within string theory can modify classical geometry to generate a horizon whose properties are  in accordance with the microscopic theory.

 To illustrate how this works out, let us analyze for simplicity the effect of the following four-derivative term in the string effective action
\begin{equation}\label{corr}
\Delta\mathcal{L} = \frac{S}{64\pi^{2}}  \left(R_{\mu\nu\a\b}R^{\mu\nu\a\b} - 4 R_{\mu\nu} R^{\mu\nu}\right) \, ,
\end{equation}
Now for the total entropy function,  instead of \eqref{entropyfinal}, one  obtains 
\be \label{efsmall}
\mathcal{E}= {\pi\over 2}  
\, ({Q^{2}\over u_{S}} + 8  u_{S}) \, .
\ee
Extremizing with respect to $u_{S}$, we obtain the attractor value of the dilaton field
\be
u_{S}^{*} =  \sqrt{Q^{2}/8} \, .
\ee
and hence the Wald entropy is given by
\be\label{macrosmall}
S_{Wald} := \mathcal{E}^{*} (Q)  := \mathcal{E}(u_{S}^{*}(Q)) = 4\pi \sqrt{Q^{2}/2} \, ,
\ee
which matches beautifully with the statistical entropy \eqref{microentropy}.

We should remember though that since the  horizon area is of order one in string units, all $\apm$ corrections are of the same order and hence the effect of all higher-derivative terms must be included at once. It turns out, however, that even  upon including the effect of all supersymmetrized F-type terms
\cite{Dabholkar:2004yr, Dabholkar:2004dq} one obtains the same results\footnote{F-type terms can be written as chiral integrals on superspace.}. 

A general scaling argument \cite{Sen:1995in} shows that up to an over all constant, the Wald entropy must have the  same form as  \eqref{macrosmall} even after all $\apm$ corrections are included up to. Moreover, by viewing the four-dimensional small black hole as an excitation of a five-dimensional black string it has been shown in
\cite{Kraus:2005vz, Kraus:2005zm} the Wald entropy is related to the coefficient of five-dimensional Chern-Simons terms. Since Chern-Simons terms are topological in nature, their coefficient is not renormalized even after including higher quantum correction. Together, these results strongly indicate that Wald entropy of small black holes upon including stringy all $\apm$ corrections  will agree with the statistical  entropy.

The agreement above and also for the entropy of quarter-BPS dyons in \S\ref{Subleading} is obtained using only the F-type terms in the string effective action. This strongly suggests a nonrenormalization theorem that  other D-terms do not renormalize the Wald entropy. For a subclass of D-type terms such a nonrenormalization theorem has recently been proven  \cite{deWit:2010za}. It would be interesting to see how it can be generalized to  all possible D-terms in this context.
\vfill\eject

\chapter{Mathematical Background \label{Mathematical}}

\section{$\mathcal{N} =4$ supersymmetry}

We summarize here some facts about the representation of the $\mathcal{N} =4$ superalgebra.  For more details see for example \cite{Kiritsis:1997gu}.

\subsubsection{Massless supermultiplets}

There are two massless representations that will be of interest to
us.
\begin{enumerate}
    \item Supergravity multiplet:\\
It contains  the metric $g_{\mu\nu}$,  six vectors $A^{(ab)}_\mu$, and  two
gravitini $\psi^a_{\mu \alpha}$.
\item Vector Multiplet:\\
It contains a vector $A_\mu$,  six scalar fields $X^{(ab)}$, and
the gaugini  $\chi^a_\alpha$, 
\end{enumerate} 
The low energy massless spectrum of a supergravity theory consists of the supergravity multiplet and  $n_v$ vector multiplets. Supersymmetry then completely fixes the form of the two derivative action. The compactification of heterotic string theory on $T^{6}$ leads to a theory in four spacetime dimensions with $\mathcal{N}=4$ supersymmetry and  $28$ abelian gauge fields which corresponds to $28-6 = 22$ vector multiplets.

\subsubsection{General BPS representations \label{BPS}}

In the rest frame of the dyon, the $\mathcal{N}=4$ supersymmetry algebra takes the form
\be
\{Q_{\a}^a, Q_{\dot{\b}}^{\dag b} \} =M  \d_{\a\dot{\b}} \, \d^{ab} \;,\;\;\;\;\; \{Q_{\a}^a, Q_{\b}^{ b} \} = \e_{\a\b} Z^{ab}\;, \;\;\;\;\;\{Q_{\dot{\a}}^{\dag a}, Q_{\dot{\b}}^{\dag b} \} = \e_{\dot{\a}\dot{\b}} \bar{Z}^{ab}
\end{equation}
where $a, b = 1, \ldots 4 $ are $SU(4)$ R-symmetry indices and $\a,\b$ are Weyl spinor indices. In a given charge sector, the central charge matrix encodes information about the charges and the moduli. To write it explicitly, we first define a central charge vector
in $\mathcal{C}^6$
\be \label{Zvector}
Z^m (\Gamma)  = \frac{1}{\sqrt{\tau_2}}(Q_R^m - \tau P_R^m)   \;, \;\;\;\; m=1,\ldots 6 \;,\;\;\;\;\;
\end{equation}
which transforms in the (complex) vector representation of $Spin(6)$. Using the equivalence $Spin(6) = SU(4)$, we can relate it to the  antisymmetric representation of $Z_{ab}$ by
\begin{equation}
Z_{ab}(\Gamma) = \frac{1}{\sqrt{\tau_2}} (Q_R - \tau P_R)^m \lambda^{m}_{ab} \;, \;\;\;\;\; m=1, \ldots 6 \label{cc}
\end{equation}
where $\lambda^{m}_{ab}$ are the Clebsch-Gordon matrices.
Since $Z(\Gamma)$ is antisymmetric, it can be brought to a block-diagonal form  by a $U(4)$ rotation
\begin{equation}\label{diag2}
\tilde{Z} = U Z U^T, \;\; U \in U(4)\;, \;\;\;\;\; \tilde{Z}_{ab} = \left(\begin{array}{c|c} Z_1 \varepsilon & 0  \\   \hline  0 & Z_2 \varepsilon \end{array}\right)\;, \;\;\; \varepsilon = \left(\begin{array}{cc} 0 & 1  \\  -1 & 0 \end{array}\right) \end{equation}
where $Z_{1}$ and  $Z_2$ are non-negative real numbers. A $U(2)$ rotation in the $12$ plane and another $U(2)$ rotation in the $34$ plane will not change the block diagonal form. Since $\varepsilon$ is the invariant tensor of $SU(2)$, the $U(2) \times U(2)$ transformation can only change independently the phases of $Z_1$ and $Z_2$. We will therefore treat more generally $Z_1$ and $Z_2$ as complex numbers.

We now split the $SU(4)$ index as $a=(r,i)$, where $r ,i=1,2$ and $i$ represents the block number.
Defining the following fermionic oscillators
\begin{equation}\label{defofA}
    \mathcal{A}_{\a}^{ i} = \frac{1}{\sqrt{2}} ( \mathcal{Q}^{1i}_\alpha+ \epsilon_{\a \b} \mathcal{Q}^{\dag \, 2 i }_\b ), \quad \mathcal{B}_{\a}^{ i} = \frac{1}{\sqrt{2}} ( \mathcal{Q}^{1i}_\alpha - \epsilon_{\a \b} \mathcal{Q}^{ \dag \, 2 i}_\b ) \;,\;\;\;\;\; \mathcal{Q}^a = U^{a}_b \,Q^b
\end{equation}
the supersymmetry algebra takes the form
\be
\{\mathcal{A}_{\dot{\a}}^{ i \dag} , \mathcal{A}_{\b}^{ j}\} = (M+ Z_i) \,\d_{\dot{\a}\b} \,\d^{ij } \;, \;\;\;\;\;\{\mathcal{B}_{\dot{\a}}^{ i \dag} , \mathcal{B}_{\b}^{ j}\} = (M- Z_i) \,\d_{\dot{\a}\b}\, \d^{ij}
\end{equation}
with all other anti-commutators being zero.

Let us conclude by giving an explicit representation for $\lambda^m_{ab}$. An  $SU(4)$ rotation which rotates the supercharges, $Q'=U Q$, acts on the Clebsch-Gordon matrices as
\be\label{transform}
U \lambda^m U^T = R^m{}_n(U) \lambda^m
\ee
where $R^m{}_n$ is an $SO(6)$ rotation matrix.
The Clebsch-Gordon matrices $\lambda^m_{ab}$ are given by the components $(C\Gamma^m)_{ab}$ where $\Gamma^m$ are the Dirac matrices of $Spin(5)$ in the Weyl basis satisfying the Clifford algebra $\{ \Gamma^m , \Gamma^n \} = 2 \delta^{mn}$, and $C$ is the charge conjugation matrix.  The Gamma matrices are given explicitly in terms of Pauli matrices by
\begin{eqnarray}
  \Gamma^1 = \sigma_1 \times \sigma_1 \times 1 &, \quad& \Gamma^4 = \sigma_2 \times 1 \times \sigma_1 \\
  \Gamma^2 = \sigma_1 \times \sigma_2 \times 1&, \quad& \Gamma^5 =  \sigma_2 \times 1 \times \sigma_2\\
  \Gamma^3 = \sigma_1 \times \sigma_3 \times 1&, \quad& \Gamma^6 = \sigma_2 \times 1 \times \sigma_3,
\end{eqnarray}
where the
The charge conjugation matrix is defined by
$C\Gamma^m C^{-1} = - {\Gamma^m}^*$
\begin{equation}\label{chargeconj}
    C = \sigma_1 \times \sigma_2 \times\sigma_2, \quad \Gamma = \sigma_3 \times 1 \times 1, \quad C\Gamma^m = \left(
                                           \begin{array}{cc}
                                             \lambda^m_{ab} & 0 \\
                                             0 & {\bar\lambda}^{m}_{{\dot a} {\dot b}} \\
                                           \end{array}
                                         \right)
\end{equation}
where the un-dotted indices transform in the spinor representation of $Spin(6)$ or the $4$ of $SU(4)$ whereas the the dotted indices transform in the conjugate spinor representation of $Spin(6)$ or the $\bar 4$ of $SU(4)$. The matrices $\lambda^m _{ab}$ thus defined have the required antisymmetry and  transform properties as in (\ref{transform}).

\section{Modular cornucopia}

We assemble here together some properties of modular forms,  Jacobi forms, and Siegel modular forms.

\subsubsection{Modular forms}
\label{modularbasic}
Let $\mathbb{H}$ be the upper half plane, \textit{i.e.}, the set of
complex numbers $\tau$ whose imaginary part satisfies $\Im ( \tau)> 0$.
Let $SL(2, \mathbb{Z})$ be the group of matrices
\begin{equation}\label{modtra}
    \left(
       \begin{array}{cc}
         a & b \\
         c & d \\
       \end{array}
     \right) \,
\end{equation}
 with integer entries such that $ad - bc =1$.

A \emph{modular form} $f (\t)$ of weight $k$ on $SL(2,\IZ)$ is a holomorphic
function on $\mathcal{H}$, that transforms as
\begin{equation}\label{modtransform0}
    f(\frac{a\t + b }{c \t + d}) = (c \t + d)^k f (\t) \, \quad \forall \quad
\left(
       \begin{array}{cc}
         a & b \\
         c & d \\
       \end{array}
     \right) \in SL(2, \mathbb{Z}) ,
\end{equation}
for an integer $k$ (necessarily even if $f(0) \neq 0$). It follows from the definition
that $f(\t)$ is periodic under $\tau \to \tau +1$ and can be written as a Fourier series
\begin{equation}\label{holmod}
  f(\t) = \sum_{n= - \infty}^{\infty} a(n) q^n \, ,  \quad q :=  e^{2 \pi i \tau} \,,
 \end{equation}
and is bounded as $\textrm{Im}(\t) \to \infty$.
If $a(0) =0$, then the modular form vanishes at infinity and is called a \emph{cusp form}.
Conversely, one may weaken the growth condition at $\infty$ to $f(\t) = \mathcal{O}(q^{-N}) $
 rather than $ \mathcal{O}(1)$ for some $N \ge 0$; then the Fourier coefficients of $f$ have the behavior
$a(n)=0$ for $n < -N$. Such a function is called a \emph{weakly holomorphic
modular form}.

The vector space over $\mathbb{C}$ of holomorphic modular forms of weight $k$ is
usually denoted by $M_k$. Similarly, the space of cusp forms of weight $k$ and
the space of weakly holomorphic modular forms of weight $k$ are denoted by $S_k$
and $M^!_k$ respectively. We thus have the inclusion
\begin{equation}\label{in}
   S_k \subset M_k \subset M^!_k \, .
\end{equation}
The growth properties of Fourier coefficients of modular forms are known:
\begin{enumerate}
\item $f \in M_{k}^{!} \Rightarrow a_{n} = \mathcal{O}(e^{C \sqrt{n}})$ as $n \to \infty$ for some $C>0$;
\item $f \in M_{k} \Rightarrow a_{n} = \mathcal{O}(n^{k-1})$ as $n \to \infty$;
\item $f \in S_{k} \Rightarrow a_{n} = \mathcal{O}(n^{k/2})$ as $n \to \infty$.
\end{enumerate}
\renewcommand{\labelenumi}{\arabic{enumi}.}

Some important modular forms on $SL(2,\IZ)$ are:
\begin{enumerate}
\item The \textit{Eisenstein series} $E_{k} \in M_{k}$ ($k \ge 4$).  The first two of these are
\begin{eqnarray}\label{eisen}
   E_4 (\t) &=& 1 + 240 \sum_{n=1}^\infty \frac{n^3q^n}{1- q^n} = 1 + 240 q + \ldots \, , \\
  E_6 (\t) &=& 1 - 504 \sum_{n=1}^\infty \frac{n^5q^n}{1- q^n} = 1 -504q + \ldots \,.
\end{eqnarray}
\item The \textit{discriminant function} $\Delta$. It is given by the product expansion
 \begin{equation}\label{discrim}
 \Delta(\tau)  =  q \prod_{n=1}^\infty {(1 - q^n)^{24}} = q - 24 q^2 + 252 q^{3} + ...
 \end{equation}
or by the formula $\Delta =  \left( E_{4}^{3} - E_{6}^{2} \right)/1728$.
\end{enumerate}
The two forms $E_4$ and $E_6$ generate the ring of modular forms, so that any
modular form of weight $k$ can be written (uniquely) as a sum of monomials
$E_4^\a E_6 ^\b$ with $4\a+ 6\b =k$.  We also have $M_{k}= \textbf{C} \cdot E_{k} \oplus
S_{k}$ and $S_{k} = \Delta \cdot M_{k-12}$, so that any $f \in M_{k}$ also has a
unique expansion as $\sum \limits_{0 \le n \le k/12} \a_{n} \, E_{k-12n} \,
\Delta^{n}$ (with $E_{0}=1$). From either representation, we see that a modular
form is uniquely determined by its weight and first few Fourier coefficients.

\subsubsection{Jacobi forms \label{Jacobi}}

Consider a holomorphic function $\varphi(\tau, z)$ from $\mathbb{H} \times
\mathbb{C}$ to $ \mathbb{C}$ which is ``modular in $\tau$ and elliptic in $z $''
in the sense that it transforms under the modular group as
\begin{equation}\label{modtransform}
   \varphi(\frac{a \t + b}{c \t + d}, \frac{z}{c\t +d}) = (c\t +d)^k \, 
e^{\frac{2\pi i m c z^2}{c\t +d}} \, \varphi(\t, z) \, , \quad \forall \quad
   \left(\begin{array}{cc}
        a&b\\
        c&d
      \end{array}
    \right) \in SL(2; \mathbb{Z})
\end{equation}
and under the translations of $z$ by $\mathbb{Z} \tau + \mathbb{Z}$ as
\begin{equation}\label{elliptic}
    \varphi(\t, z + \lambda \tau + \mu) = e^{-2\pi i m(\lambda^2 \t
+ 2 \lambda z)}
\varphi(\t, z), \quad
\forall
\quad \l, \m \in \mathbb{Z} \, ,
\end{equation}
where $k$ is an integer and $m$ is a positive integer.

These equations include the periodicities $\varphi(\tau+1,z) = \varphi(\t,z)$ and
$ \varphi(\tau,z+1) = \varphi(\t,z)$, so $\varphi$ has a Fourier expansion
\begin{equation}\label{fourierjacobi}
    \varphi(\t, z) = \sum_{n, r} c(n, r) \, q^n \, y^r \ , \qquad \qquad (q :=
e^{2\pi i \t}, \; y := e^{2 \pi i z}) \ .
\end{equation}
Equation \eqref{elliptic} is then equivalent to the periodicity property
\be\label{cnrprop}
c(n, r) = C(4 n m - r^{2} ; \, r) \ ,
\qquad \mbox{where} \; C(d;r) \; \mbox{depends only on} \; r \, (\mod\, 2m) \ .
\ee

The function $\varphi(\tau, z)$ is called a \emph{holomorphic Jacobi form} (or
simply a \emph{Jacobi form}) of weight $k$ and index $m$ if the coefficients
$C(d;r)$ vanish for $d<0$, \textit{i.e.} if 
\begin{equation}\label{holjacobi}
    c(n, r) = 0 \qquad \textrm{unless} \qquad 4mn \ge r^2 \ . 
    \end{equation}
It is called a \emph{Jacobi cusp form} if it satisfies the stronger condition that
$C(d;r)$ vanishes unless $d$ is strictly positive, \textit{i.e.}
\begin{equation}\label{cuspjacobi}
    c(n, r) = 0 \qquad \textrm{unless} \qquad 4mn > r^2 \ , 
\end{equation}
and conversely, it is called a \emph{weak Jacobi form} if it satisfies the weaker condition
\begin{equation}\label{weakjacobi}
    c(n, r) =  0\qquad   \textrm{unless}  \qquad n \geq 0 \, 
\end{equation}
rather than \eqref{holjacobi}. 

\subsubsection{Theta functions}

In this section, we collect definitions and useful properties of
theta function. The Jacobi theta function is defined by
\be
\th[^a_b](v|\t)=\sum_{n\in \IZ}q^{{1\over 2}\left(n-a\right)^2}
e^{2\pi i\left(v-b\right)\left(n-a\right)}
\,,\label{t1}\ee
where $a,b$ are real and $q=e^{2\pi i\t}$. It satisfies the modular properties
\bea
\th[^a_b](v|\t+1)&=&e^{-i\pi a(a-1)}\th[^a_{a+b-\frac12}](v|\t) \\
\th[^a_b]\left(\frac{v}{\t}|-\frac{1}{\t}\right)&=&
e^{2i\pi ab + i \pi \frac{v^2}{\t}}
\th[^a_{b}](v|\t)
\eea
The Jacobi-Erderlyi theta functions are the values at half periods,
\be
\th_1(z|\tau)=\th[^\half_\half](z|\t),\quad
\th_2(z|\tau)=\th[^\half_0](z|\t),\quad
\th_3(z|\tau)=\th[^0_0](z|\t),\quad
\th_4(z|\tau)=\th[^0_\half](z|\t)
\ee
In particular,
\be
\label{modt1}
\theta_1(v/\tau,-1/\tau) = i \sqrt{-i \tau}  e^{i\pi  v^2/\tau}
\theta_1(v,\tau)
\ee
The Dedekind $\eta$ function is defined as
\be
\eta(\t)=q^{1\over 24}\prod_{n=1}^{\infty}(1-q^n)
\,.\label{t10}
\ee
It satisfies the modular property
\be
\label{modde}
\eta\left( - \frac{1}{\t} \right) = \sqrt{-i\t} \eta(\tau)
\ee
It is related to the Jacobi-Erderlyi theta functions by the identities
\bea
{\partial\over \partial v}\th_1(v)|_{v=0} &= & 2\pi{}~\eta^3(\t)
\label{t11} \\
\th_2(0|\t)\th_3(0|\t)\th_4(0|\t)&=&2\eta^3
\label{t13}
\eea
The partition function of a single left-moving boson is given by 
\begin{equation}
Z_{boson}(\tau) :=  \textrm{Tr} (q ^{L_{0}}) = \frac{1}{\eta(\tau)} \, .
\end{equation}

\subsubsection{Siegel modular forms}
Let $Sp(2, \mathbb{Z})$ be the group of $(4\times 4)$ matrices $g$ with
integer entries satisfying $g J g^t =J$ where
\begin{equation}
 \quad J \equiv \left(
                                  \begin{array}{cc}
                                    0 & -I_2 \\
                                    I_2 & 0 \\
                                  \end{array}
                                \right)
\end{equation}
is the symplectic form. We can write the element $g$ in block form as
\begin{equation}\label{sp}
   \left(
  \begin{array}{cc}
    A & B \\
    C & D \\
  \end{array} \right),
  \end{equation}
where $A, B, C, D$ are all $(2\times 2)$ matrices with integer entries.
Then the condition $g J g^t =J$ implies
\begin{equation}\label{cond}
   AB^t=BA^t, \qquad CD^t=DC^t, \qquad AD^t-BC^t= \mathbf{1}\, ,
\end{equation}
Let $\mathbb{H}_2$ be the (genus two) Siegel upper half plane, defined as the set
of $(2\times 2)$ symmetric matrix $\Omega$ with complex entries
\begin{equation}\label{period}
   \Omega = \left(
              \begin{array}{cc}
                \t & z \\
                z & \sigma \\
              \end{array}
            \right)
\end{equation}
satisfying
\begin{equation}\label{cond1}
   \textrm{Im} (\t) > 0, \quad \textrm{Im} (\sigma) > 0, \quad
   \det (\Im(\O)) > 0 \, .
\end{equation}
An element $g \in Sp(2, \mathbb{Z})$ of the form (\ref{sp}) has a natural action on
$ \mathbb{H}_2$ under which it is stable:
\begin{equation}\label{trans}
    \Omega \to (A \Omega + B )(C\Omega + D ) ^{-1}.
\end{equation}
The matrix $\Omega$ can be thought of as the period matrix of a genus two 
Riemann surface\footnote{See \cite{Gaiotto:2005hc, Dabholkar:2006xa, Banerjee:2008yu} for a
discussion of the connection with genus-two Riemann surfaces.} on which there is
a natural symplectic action of $Sp(2, \mathbb{Z})$.

A Siegel form $F(\Omega)$ of weight $k$ is a holomorphic function $ \mathbb{H}_2
\rightarrow \mathbb{C}$ satisfying
\begin{equation}\label{trans2}
    F [(A \Omega + B )(C\Omega + D ) ^{-1}] =  \{\det{(C\Omega
+ D )}\}^k
    F (\Omega).
\end{equation}
A Siegel modular form can be written in terms of its Fourier series
\begin{equation}\label{siegelfourier}
    F(\Omega) = \sum a(n,r, m)  \, q^n y^r  p^m \, .
\end{equation}

The Siegel modular form which makes its appearance in the present physics problem of counting $\mathcal{N}=4$ dyons  is the Igusa  form $\Phi_{10}$ 
which is the unique (cusp) form\footnote{It is a `cusp' form because it vanishes at `cusps' which correspond to $z=0$ and its images.} of weight $10$. This Siegel modular form is a very interesting mathematical object and has a number of useful properties directly relevant for the present physical application. In particular, it can be constructed very explicitly in two different ways in terms of familiar modular forms and theta functions by using two different `lifts.' These constructions are called lifts because they allow us to construct the  Igusa cusp form which  is a function of three variables using the Fourier expansions of a weak Jacobi forms which are  functions of only two variables.

\begin{itemize}
\item
\textit{Additive lift}\\
Consier the function   $ \psi(\t, z)$ 
\begin{equation}\label{phi10}
    \psi(\t, z) = \eta^{18}(\t) \, {\vartheta_1^2(\t, z)} \, .
\end{equation}
which is a weak Jacobi form of weight $1$ and index $10$ (see \S\ref{Jacobi}  for definitions). It admits a Fourier expansion
\begin{equation}
\psi(\t, z) = \sum_{n, r} c_{{10}}(n, r) q^{n} y^{r} \qquad  q:= e^{2\pi i \t} \, y:= e^{2\pi i z} \, .
\end{equation}
{}From the properties of weak Jacobi forms, it follows that the Fourier coefficients 
$c_{10}(n, r)$ depend only on the combination $4n-r^{2}$ and hence we can write
 $c_{10}(n, r) = C_{10}(4n-r^{2})$ for some function $C_{10}$.
The additive lift then gives the Fourier expansion of the Igusa cusp form  in terms of the Fourier coefficients of $\psi(\tau, z)$ as 
\begin{equation}\label{igusa-additive}
    \Phi_{10}(\Omega) = \sum_{n, m, l} a(m, n, l) p^m q^n y^l \, , \quad p:=e^{2\pi i \sigma} \, , 
\end{equation}
where $a(m, n, l)$ are defined by 
\be\label{Cnrm} 
a(n, r, m) = \sum_{{d | (n,r,m)}\atop{d \ge 1}} d^{k-1} C_{10}(\frac{4 m
n - r^{2}}{d^{2}}) 
\ee 
This lift is  `additive' in that it gives   a sum representation of the Igusa form. 
\item \textit{Multiplicative lift} \\
Consider the function  $\chi (\t, z)$
\begin{equation}\label{phizero}
    \chi (\t, z) = 8 \left( \frac{\vartheta_2(\t, z)^2}{\vartheta_2(\t)^2} +
    \frac{\vartheta_3(\t, z)^2}{\vartheta_3(\t)^2} +
    \frac{\vartheta_4(\t, z)^2}{\vartheta_4(\t)^2} \right) \, ,
\end{equation}
which is weak Jacobi form of weight $0$ and index $1$ 
with a Fourier expansion
\begin{equation}\label{phizeroFE}
\chi(\t, z) = \sum_{n, r} c_{{0}}(n, l) q^{n} y^{l} \qquad  q:= e^{2\pi i \t} \,\, , \, y:= e^{2\pi i z} \, .
\end{equation}
This function arises in physics applications as the elliptic genus of the $K3$ surface (see appendix \eqref{K3} for details).
{}Once again,
$c_{0}(n, l)$ depend only on the combination $d:=4n-l^{2}$ and hence we can write
\begin{equation}\label{defC}
c_{0}(n, l) = C_{0}(4n-l^{2}) 
\end{equation}
which defines the function $C_{0}(d)$.
The multiplicative lift gives a product representation of the Igusa cusp form in terms of $C_{0}(d)$:
\begin{equation} \label{final2}
  \Phi_{10}(\Omega) = p q y \prod_{(s, t, r) >0}( 1 - p^s q^t y^r)^{C_0(4st-r^2)},
\end{equation}
in terms of $C_0$ given by (\ref{phizero}, \ref{phizeroFE}).
Here the notation $(s, t, r) >0$ means that either $s >0, t, r\in
\mathbb{Z}$, or $s=0, t >0, r\in \mathbb{Z}$, or $s = t = 0, r < 0$.

This lift is  `multiplicative' in that it gives  a product representation of the Igusa form. 
\end{itemize}

\section{ A few facts about $K3$ \label{K3}}

\subsubsection{K3 as an orbifold}

``Kummer's third surface''  or K3 has played an important
role  in many developments concerning duality.
Let us recall some of its properties.  $ K3$ is  a four dimensional
 manifold which has
$SU(2)$ holonomy. To understand what this means, consider a generic 4d
real manifold. If you take a vector in the tangent space at point $P$,
parallel transport it, and come back to point $P$, then, in general, it
will be rotated by an $SO(4)$ matrix:
\bea
V_i (P) \rightarrow O_{i j }\ V_i (P) \ \ \ \ O_{ij} \in SO(4).
\eea
Such a manifold  is then said to
to have  $SO(4)$ holonomy.
In the case of K3, the holonomy is a subgroup of $SO(4)$,
namely $SU(2)$. The smaller the holonomy group, the more
``symmetric'' the space. For example, for a torus, the holonomy group
consists of  just the identity because the space is flat and
Riemann curvature is zero;
so, upon parallel transport along a closed loop, a vector
comes back to itself.
For a K3, there $\it{is}$ nonzero curvature
but it is not completely arbitrary: the Riemann
tensor is non-vanishing but the Ricci
tensor $R_{ij}$ vanishes. Therefore, K3 can alternatively
be defined as the manifold of compactification that solves
the vacuum Einstein equations.

Only other thing about K3 that we need to know
is the topological information.  A surface can have nontrivial cycles
which cannot be shrunk to a point. For example, a torus
has two nontrivial 1-cycles.
The number of nontrivial k-cycles
which cannot be smoothly deformed into each other  is given by
the $ k$-th Betti number $ b_k$ of the surface.
The number of non-trivial $ k$-cycles is in one to one correspondence
with the number of harmonic $ k$-forms on the surface
given by the $ k$-th
de-Rham cohomology \cite{Green:1987sp, Green:1987mn}.
A harmonic k-form $ F_k$ satisfies the Laplace equation, or equivalently
satisfies the equations
\bea
d^\ast F_k=0,\qquad d F_k =0
\eea
A manifold always has a harmonic 0-form, {\it viz.}, a constant, and a
harmonic 4-form, {\it viz.},  the volume from, assuming we can integrate on it.
K3 has no harmonic 1-forms or  3-forms, but has
 22 harmonic 2-forms.
So, the Betti numbers for K3 are:
\bea
b_0 =1, \qquad b_1 =0,\qquad b_2 =22,\qquad b_3=0,\qquad b_4=1.
\eea
Out of the $ 22$ 2-forms, $19$ are anti-self-dual,
and $ 3$ are self-dual. In other words,
\bea
b_2^s =3,\qquad b_2^a=19.
\eea
This is all the information one
needs to compute the massless spectrum of compactifications
on K3.

K3 has a simple description as a $ \bZ_2$ orbifold of
a 4-torus. Let $(x_1, x_2, x_3, x_4)$
be the real coordinates of the torus  $ \bT^4$. Let us further
take the torus to be a product
$ \bT^4 = \bT^2\times \bT^2$. Let us introduce complex coordinates
$(z_1, z_2)$, $z_1 = x_1 + ix_2$ and  $z_2 = x_3
+ ix_4$. The 2-torus with coordinate $z_1$ is defined
by the identifications $z_1 \sim z_1 +1 \sim z_1 +i$, and similarly
for the other torus.
The tangent space group is
$Spin(4) \equiv SU(2)_1 \times SU(2)_2,$
and the vector representation is ${ \bf  4v} \equiv ( {\bf 2, 2} )$.
If we take a subgroup
$
SU(2)_1 \times U(1)
$
of $ Spin(4)$, then the vector decomposes as
\begin{eqnarray}
{\bf 4v} = {\bf 2_+} \oplus {\bf \bar 2_-}.
\eea
The coordinates$
\left(
  z_1,   z_2 
\right)
$
 transform
as the doublet  ${\bf 2_+}$
and $\left({\bar z_1, \bar z_2}\right)$ as
the  $\bf \bar 2_- $.
The $ \bZ_2 =\{ 1, I\}$ is generated by
\bea
I : (z_1, z_2) \rightarrow (-z_1, -z_2).
\eea
This $\bZ_2$
is a subgroup and in fact the center of $SU(2)_1$.
Consequently, as we shall see, the resulting manifold has
$ SU(2)$, indeed a $ \bZ_2$ holonomy. For a torus coordinatized
by $ z_1$,
there are 4 fixed points of $z_1 \rightarrow -z_1$
Altogether, on $ \bT^4/\bZ_2$,
there are 16 fixed points.

Let us calculate the number of harmonic forms on
this orbifold. To begin with, we have on the
torus $ \bT^4$, the following harmonic forms:
\begin{eqnarray}
1&& 1\nonumber\\
4 && dx^i \ \ \ \nonumber \\
6 && dx^i \wedge dx^j \nonumber \\
4 && dx^i \wedge dx^j \wedge dx^l\nonumber \\
1 && dx^i \wedge dx^j \wedge dx^k \wedge dx^l.
\end{eqnarray}
The first column gives the number of forms indicated in the second column
where the indices
 $ i, j, k, l$ take values $ 1,\cdots 4$.
Under the reflection $ I$, only the even forms
$ 1, dx^i\wedge dx^j$, and $dx^i \wedge dx^j \wedge dx^k \wedge dx^l$ survive.
\bea
\begin{tabular}{llllll}
0-form & 1 & & 1 && \\
1 & 4 && 0 && \\
2 & 6 & $ { {1+ I \over 2} \atop {\longrightarrow}}$ &6 & & \\
3 & 4 && 0 && \\
4 & 1 && 1 && \\
\end{tabular},
\eea
where the  second column give the number of forms on the torus
and the third column the number of forms that survive the projection.
Let  us look at the 2-forms  from the torus that survive the $ \bZ_2$ projection.
By taking the combinations
$$
dx^i \wedge dx^j  \pm {1 \over 2} \epsilon^{ijkl} dx^k \wedge dx^l
$$
we see that three of these 2-forms are self-dual and the remaining three
are anti-self-dual.

At the fixed point of the orbifold symmetry there is a curvature singularity.
The singularity can be repaired as follows.
We cut out a ball of radius $ R $ around each point, which has a boundary $S^3/{\bZ_2}$,
 replace it with a noncompact smooth manifold that is also Ricci flat
and has a boundary $S^3/{\bZ_2}$,
and then take the limit $ R\rightarrow 0$.
The required noncompact  Ricci-flat manifold  with
boundary $S^3/{\bZ_2}$ is known to exist and
is called the Eguchi-Hanson space. The Betti number
of the Eguchi Hanson space are
$b_0=b_4=1$ ad $ b_2^a =1$.
Therefore, each fixed point contributes an anti-self-dual 2-form
which corresponds to a nontrivial 2-cycle in the Eguchi-Hanson
space that would be  stuck at the fixed point in the limit $ R\rightarrow 0$.

Altogether, we get $b_0=1, b_2^s =3$, $ b_2^a=3 + 16=19, b_4=1$,
and $ b_1=b_3=0$  giving us the cohomology of K3. It obviously has
$SU(2)$ holonomy. Away from the fixed point, a parallel transported
vector goes back to itself, because all the curvature is concentrated
at the fixed points. As we go around the fixed point a vector is returned
to its reflected image (for instance, $(dz_1, dz_2) \rightarrow -(dz_1, dz_2)$),
\textit{i. e.}, transformed by an element of $ SU(2)$.

In string theory there is no need to repair the singularity by hand.
We shall see in $ \S{5.3}$ and $\S{5.4}$ that the twisted states
in the spectrum of Type-II string moving on an orbifold
automatically take care of the repairing.
The twisted states  somehow know
about  the Eguchi-Hanson manifold that would be necessary
to geometrically repair  the singularity.

\subsubsection{Elliptic genus of $K3$}
Consider a 
two-dimensional superconformal field theories (SCFT) with $(2, 2)$ or more
worldsheet supersymmetry\footnote{An SCFT with $(r, s) $ supersymmetries has $r$
left- moving and $s$ right-moving supersymmetries.}.  We denote the
superconformal field theory by $\sigma(\CM)$ when it corresponds to a sigma model
with a target manifold $\mathcal{M}$.  Let $H$ be the Hamiltonian in the Ramond
sector, and $J$ be the left- moving $U(1)$ R-charge. The elliptic genus
$\chi(\tau, z; \mathcal{M})$ is then defined  \cite{Witten:1986bf, 
Alvarez:1987wg, Ochanine:1987} as a trace over the Hilbert space
$\mathcal{H}_R$ in the Ramond sector
\begin{equation}\label{ell}
   \chi(\tau, z; \mathcal{M}) = { \tr{\mathcal{H}_R}} \left( q^H y^J
(-1)^F \right) \, .
\end{equation}
where $F$ is the fermion number.
An elliptic genus so defined satisfies the modular transformation property
(\ref{modtransform}) as a consequence of modular invariance of the path integral.
Similarly, it satisfies the elliptic transformation property (\ref{elliptic}) as
a consequence of spectral flow.  Furthermore, in a unitary SCFT, the positivity
of the Hamiltonian implies that the elliptic genus is a weak Jacobi form.

A particularly useful example in the present context is $ \sigma(K3)$, which is a
$(4, 4)$ SCFT whose target space is a $K3$ surface. The elliptic genus is a
topological invariant and is independent of the moduli of the $K3$. Hence, it can
be computed at some convenient point in the $K3$ moduli space, for example, at
the orbifold point where the $K3$ is the Kummer surface. At this point, the
$\sigma(K3)$ SCFT can be regarded as a $\mathbb{Z}_2$ orbifold of the $
\sigma(T^4)$ SCFT which is an SCFT with a torus $T^4$ as the target space. A
simple computation using standard techniques of orbifold conformal field theory
yields \cite{Ginsparg:1988ui} the formula for the elliptic genus we introduced earlier in \eqref{phizero}:
\begin{equation}\label{phizero}
    \chi (\t, z) = 8 \left( \frac{\vartheta_2(\t, z)^2}{\vartheta_2(\t)^2} +
    \frac{\vartheta_3(\t, z)^2}{\vartheta_3(\t)^2} +
    \frac{\vartheta_4(\t, z)^2}{\vartheta_4(\t)^2} \right) \, .
\end{equation}
The first term can be seen to arise from the untwisted projected partition function, the second from the twisted, unprojected partition function and the third from the twisted, projected partition function. 

Note that for $z=0$, the trace (\ref{ell}) reduces to
the Witten index of the SCFT and correspondingly the elliptic genus reduces to
the Euler character of the target space manifold. In our case, one can readily
verify from (\ref{K3}) and (\ref{phizero}) that $\chi(\t, 0; K3)$ equals $24$
which is the Euler character of $K3$.

\bibliography{final}
\bibliographystyle{JHEP}

\end{document}